\documentclass[onecolumn,authoryear]{els-mrw} 

\usepackage{amsmath,amssymb,amsfonts,amsthm,makeidx,graphicx}
\usepackage{natbib}
\usepackage{soul}
\usepackage{txfonts}
\usepackage{helvet}
\usepackage{ulem}

\newcommand{\MJup}{\ensuremath{M_{\mathrm{Jup}}}}

\newcommand{\MSun}{\ensuremath{M_{\odot}}}

\newcommand{\mic}{$\mu$m}

\newcommand\farcs{\hbox{$.\!\!^{\prime\prime}$}}
\newcommand{\aap}{Astron. \& Astrophys.}
\newcommand{\mnras}{Mon. Not. Roy. Astron. Soc.}
\newcommand{\aj}{Astron. J.}
\newcommand{\apj}{Astrophys. J.}
\newcommand{\apjl}{Astrophys. J. Lett.}
\newcommand{\apjs}{Astrophys. J. Suppl.}
\newcommand{\araa}{Ann. Rev. Astron. Astrophys.}

\newcommand{\icarus}{Icarus}
\newcommand{\pasp}{Pub. Astron. Soc. Pacific}

\newcommand{\nat}{Nature}

\newcommand{\pasa}{Publications of the Astronomical Society of Australia}
\newcommand{\ao}{AO}

\begin{document}

\chapter{Direct imaging of exoplanets}\label{chap1}

\author[1,2]{Alice Zurlo}%

\address[1]{Instituto de Estudios Astrof\'isicos, Facultad de Ingenier\'ia y Ciencias, Universidad Diego Portales, Av. Ej\'ercito Libertador 441, Santiago, Chile}, 
\address[2]{Millennium Nucleus on Young Exoplanets and their Moons (YEMS)}

\articletag{Chapter Article tagline: update of previous edition, reprint..}

\maketitle




\begin{glossary}[Nomenclature]
\begin{tabular}{@{}lp{34pc}@{}}
4QPM & Four quadrant phase-mask \\
ADI & Angular Differential Imaging \\
ALC & Apodized Lyot Coronagraph \\
AO & Adaptive Optics \\
APP & Apodizing phase plate \\
au & Astronomical Unit \\
BD & Brown Dwarf \\
CA & Core Accretion \\
DM & Deformable Mirror \\
eELT & European Extremely Large Telescope \\
ERIS & Enhanced Resolution Imager and Spectrograph \\
FOV & Field Of View  \\
GI & Gravitational Instability \\
HCI & High-Contrast Imaging \\
JWST & James Webb Space Telescope \\
IWA & Inner Working Angle \\
LOCI & Locally Optimized Combination of Images \\
NACO & NAos COnica \\
NIR & Near-Infrared \\
OWA & Outer Working Angle \\
PCS & Planetary Camera and Spectrograph \\
PSF & Point Spread Function \\
RDI & Reference Differential Imaging \\
RV & Radial Velocity\\
SDI & Spectral Differential Imaging \\
SNR & Signal-to-Noise Ratio\\
SPHERE & Spectro-Polarimetric High-contrast Exoplanet REsearch \\
SR & Strehl Ratio \\
VLT & Very Large Telescope \\
VLTI & Very Large Telescope Interferometer \\
\end{tabular}
\end{glossary}

\begin{glossary}[Glossary]

{\bf Adaptive optics systems}: Adaptive optics systems enhance ground-based telescopes' resolution by correcting atmospheric turbulence, enabling clearer observations of celestial objects.

{\bf Airy Disk}: In optics, particularly in telescopes, the Airy disk refers to the central bright disk surrounded by concentric rings in the image of a point source of light. It is a diffraction pattern caused by the circular aperture of the telescope. \\

{\bf Apodizer}: An apodizer is an optical element used in telescopes or imaging systems to modify the intensity distribution of light across the aperture. It can be used to reduce the intensity of diffraction patterns or to improve image contrast by altering the shape of the point spread function. 

{\bf Attenuate}: In astronomy, attenuate refers to the reduction in the intensity of radiation or light. \\

{\bf Cold start}: In the context of planetary formation, a cold start refers to a scenario where a planet forms via gravitational collapse from a cold, dense cloud of gas and dust without undergoing a phase of significant heating.

{\bf Hot start}: In contrast to a cold start, a hot start in planetary formation refers to a scenario where a planet forms through rapid accretion of material often involving significant heating processes.

{\bf Inner Working Angle (IWA)}: In astronomical observations, the inner working angle refers to the separation where the peak flux of the star is attenuated by 50\%.

{\bf Lyot stop}: A Lyot stop is a specialized aperture stop used in certain types of coronagraphs to block the light of the central star, particularly stray light and diffraction artifacts, while allowing the light from the target object to pass through. It helps to improve the contrast of faint objects near bright sources.

{\bf Occulting Mask}: An occulting mask is a device used in telescopes or coronagraphs to block or mask out the light from a central bright source, such as a star, allowing fainter objects nearby, such as exoplanets or protoplanetary disks, to be observed more easily.

{\bf Outer Working Angle (OWA)}: Similar to the Inner Working Angle (IWA), the outer working angle refers to the maximum angular separation at which the adaptive optic of the system is performing.

{\bf Point Spread Function (PSF)}: The point spread function describes the response of an imaging system, such as a telescope or camera, to a point source of light. It characterizes how the light from a single point in the object is spread out in the resulting image.

{\bf Post-processing}: Post-processing in astronomy refers to the digital manipulation and enhancement of images or data obtained from telescopes or other observation instruments after they have been initially captured. This can involve techniques such as noise reduction, image stacking, and enhancement of specific features.

{\bf Snow line}: Also known as the frost line or ice line, the snow line is the distance from a central star at which the temperature in a protoplanetary disk drops low enough for volatile compounds like water, methane, and ammonia to condense into solid ice grains, affecting the composition of forming planets.

{\bf  Speckle}: In astronomy, speckle refers to small-scale fluctuations in brightness or intensity observed in images of celestial objects caused by atmospheric turbulence or imperfections in optical systems.

{\bf Strehl Ratio}: The Strehl ratio is a measure of the quality of an optical system, such as a telescope or camera. It compares the peak intensity of the point spread function (PSF) of the system to the ideal PSF for a perfect optical system. A higher Strehl ratio indicates better image quality and reduced aberrations.

{\bf Transitional Disk}: In astronomy, a transitional disk refers to a type of protoplanetary disk surrounding a young star that exhibits a gap or inner hole in its dust and gas distribution. These disks are believed to be in a transitional phase of planetary formation, possibly indicating the presence of newly forming planets that have cleared material from their orbits.

{\bf  Warm start}: A warm start, similar to a hot start, describes a scenario in planetary formation where a planet forms through the rapid accretion of material onto a pre-existing core, but without undergoing as much heating as in a typical hot start scenario. This can result in different planetary compositions and structures.

\end{glossary}

\begin{abstract}[Abstract]
Over the past four decades, the exploration of planets beyond our solar system has yielded the discovery of over 5600 exoplanets orbiting different stars. Continuous advancements in instrumentation and cutting-edge techniques empower astronomers to unveil and characterize new exoworlds with increasing frequency. Notably, direct imaging, also called high-contrast imaging, stands out as the only method capable of capturing photons emitted directly from the planetary bodies.

This innovative technique proves particularly advantageous for scrutinizing nascent planetary systems, where planets shine brilliantly and emit significant heat during their initial developmental phases. Direct imaging provides comprehensive visuals of the entire system, encompassing the central star, potential circumstellar disks, and any additional companions.

However, the complexity of imaging an object millions of times fainter than its parent star necessitates state-of-the-art instrumentation. High-contrast imaging demands cutting-edge tools such as extreme adaptive optics systems, telescopes exceeding 8 meters in diameter, coronagraphs, and modern imagers. The pivotal role of post-processing cannot be overstated in the quest for detecting and characterizing planets through direct imaging. Substantial progress has been made in this realm since the first detection in 2005.

This method has not only facilitated the discovery of numerous planets but has also presented invaluable opportunities to explore the properties of young substellar companions, both planets and brown dwarfs. Insights into their interactions with parent disks or other companions within the system, the composition of their atmospheres, and the identification of still accreting planets, also known as "protoplanets," contribute significantly to our understanding of planet formation scenarios. The continued refinement of direct imaging techniques promises to unveil further revelations in the captivating field of exoplanetary exploration.

\end{abstract}

\section{Objectives}

\begin{itemize}
    \item Learn the differences between the exoplanet detection techniques;
    \item Understand the effects of the atmosphere on the observations;
    \item Learn about the instrumentation needed to perform high-contrast imaging;
    \item Know the possible post-processing techniques of the high-contrast imaging;
    \item Investigate the history of the direct imaging technique and some of the benchmark results;
    \item Learn about the different theories of planet formation;
    \item Understand what are the protoplanets, and why we are so interested in them;
    \item Discern the different types of sub-stellar objects.
    
\end{itemize}

\section{Introduction}\label{chap1:sec1}

\begin{figure}[t]
  \centering
    \includegraphics[width=0.9\textwidth]{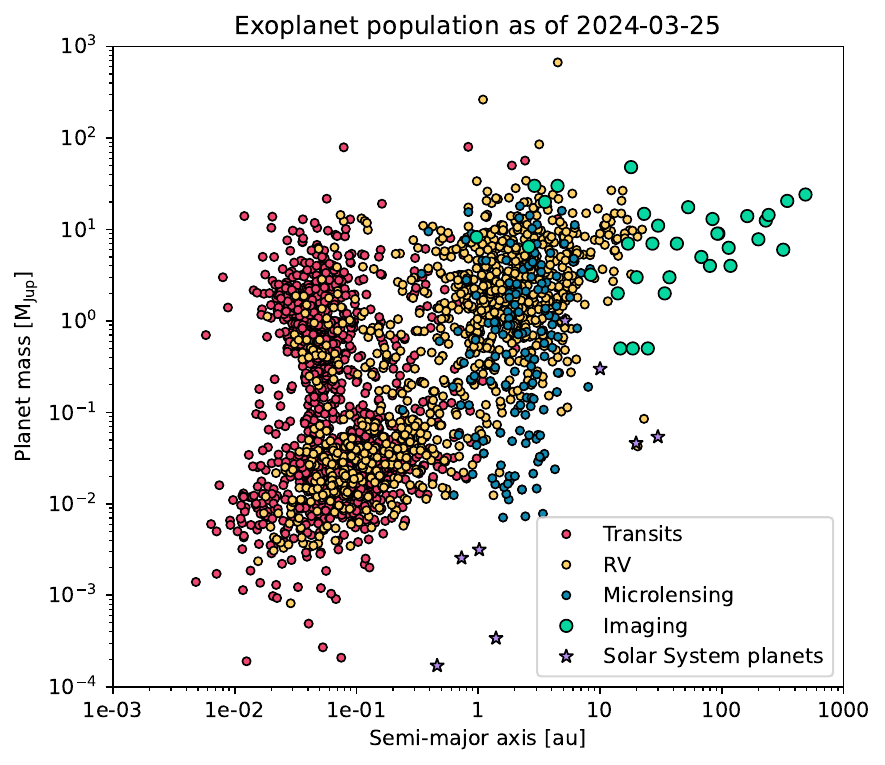}
  \caption{Exoplanet population in the mass (in Jupiter masses) vs separation (in au) diagram. Each technique is represented with a different color and populates a different region of the diagram. Direct imaging detects planets at wide separations, exploring the outer part of planetary systems.}
  \label{f:exo}
\end{figure}

In our Galaxy, as of February 2024, there are currently $\sim$5600 known exoplanets, which are planets orbiting stars other than the Sun\footnote{https://exoplanet.eu/catalog/}. Different detection techniques have contributed to the diverse sample of exoplanets, each focusing on specific types of objects. Transits, {the slight dimming of a star's light as an exoplanet passes in front of it from our point of view,} reveal planets in close proximity to their parent stars and account for the majority of current detections, spanning a wide range of masses. Radial velocities, { the wobble in a star's motion caused by the gravitational pull of an orbiting planet,} favor short separations and can detect low-mass planets. On the other hand, gravitational microlensing, {the brief brightening of a distant star due to the gravitational field of a foreground star-planet system acting as a magnifying lens,} is sensitive to planets at intermediate separations but lacks follow-up observations. Additionally, astrometry, { the measuring of tiny shifts in a star's position caused by the gravitational pull of an orbiting planet,} can detect massive planets at intermediate separations. Figure~\ref{f:exo} illustrates the mass vs separation diagram, displaying the current exoplanet population.

In addition to these techniques, direct imaging, also called high-contrast imaging (HCI) stands out as the only method capable of directly observing giant gaseous exoplanets at wide separations, particularly in young systems (less than 500 Myr). Indirect techniques, in contrast, are influenced by stellar activity and primarily target older systems. 

Direct imaging is the only technique exploring the outer part of planetary systems (Fig.~\ref{f:exo}), and the one that gives insights into giant planets still forming or recently formed. This technique started to detect planetary mass companions in 2005, and since then it has detected over 30 planets and several brown-dwarf companions. Even if the number of detections is significantly lower than other indirect techniques, direct imaging has several advantages that make every detection key to better understanding planet formation and dynamics of planetary systems. 

Direct imaging consists of masking the light of the host star with a coronagraph or a nulling technique to reveal what there is around it. It is a challenging technique, as the contrast, i.e. the brightness ratio between the planet and its host star, is of the order of 10$^{-6}$ even in rather favorable cases. For comparison, it is similar to detect a coin of the size of a dime close to a lighthouse, at a distance of 1 km. Also, the residual light of the star is brighter than the object of interest. The residual light is also called ``the speckle pattern'', produced by the imperfections of the optical elements that scatter the light of the central star, it resembles a halo of thousands of bright point-spread functions. Advanced extreme adaptive optics (AO) instrumentation, coupled with coronagraphy and state-of-the-art imagers, are crucial to attenuate the speckle pattern and reveal planets detected through direct imaging, as presented in details in Sec.~\ref{s:ins}. 

To detect faint companions, buried under the speckle pattern, and sometimes very close to the central star, the post-processing phase is key. The post-processing methods are intrinsically connected to the observing technique. The most common techniques optimized to detect and characterize companions are spectral differential imaging (SDI), angular differential imaging (ADI), and reference differential imaging (RDI). Those post-processing techniques are presented in Sec.~\ref{s:post}.  

Because of the complexity of the method, the number of the systems with confirmed HCI companions is significantly lower than the other indirect methods. The most emblematic direct imaging systems observed and studied so far are presented in Sec.~\ref{s:di_systems}. Some benchmark systems are described in more detail in the dedicated subsections. Every system directly detected is unique and provides insights into the formation scenario, the dynamics of the planet(s) in the system, the disk-to-planet interaction, and the composition of the atmosphere of the companion itself. 

This powerful technique focuses on very young systems, as during the first stages of planet formation the entropy is high and planets are hot, self-luminous, and still very bright. Direct imaging is the only technique able to spatially separate the photons of young planets and it permits the measurement of their spectra. The technique uniquely paints a picture of the planetary system just after its birth, giving clues to the study of planet formation mechanisms (presented in Sec.~\ref{s:form}).

Planets that are still forming, also known as ``protoplanets'' have been imaged with high-contrast imaging. Those objects are still undergoing mass accretion and they are an invaluable insight to understand where and how planets form. I present the known protoplanets and the challenging hunt for this kind of object in Sec.~\ref{s:proto}.

The objects of interest of the high-contrast imaging technique are not only planets but in general all sub-stellar objects, including brown dwarf companions. I present the different spectra of the objects of interest of the direct imaging technique in Sec.~\ref{s:atm}. The conclusions of this chapter and the future prospects of the direct imaging technique are presented in Sec.~\ref{s:conc}.

\section{Instrumentation}
\label{s:ins}

The principles of adaptive optics are to attenuate the influence of the atmosphere by measuring its behavior at a high-frequency rate and correcting for its contribution quasi-simultaneously. The light of the stars, which by approximation comes from an infinite distance, should be a perfect plane wave when entering the pupil of the telescope. Following the law of the diffraction by a circular aperture, the angular resolution of a telescope with an aperture size $D$ is defined as:
\begin{equation}
\theta_{\rm diff} = 1.22 \frac{\lambda}{D},
\end{equation}    
where $\lambda$ is the wavelength of the observation. $\theta_{\rm diff}$ is the radius of the inner Airy disk, of the order of 30 mas for $\lambda=1$ $\mu$m and $D \sim 8$ m.  What is observed on the focal plane of a telescope is essentially the Fourier transform of the wavefront impinging on the pupil plane multiplied by the pupil shape \citep[Fraunhofer approximation: see, e.g.,][]{2009ARA&A..47..253O}. The intensity of the Airy function is mostly concentrated in the first peak, inside $\theta_{\rm diff}$, then has secondary rings separated by minima of the function, with decreasing energy.    

For ground-based telescopes, the light from the stars passes through the Earth's atmosphere. This latter is formed by layers of turbulent ``bubbles'', of size $r_0$ (Fried parameter), within which the wavefront remains coherent. The bubbles have different refraction coefficients so that the wavefront is corrugated before entering the pupil of the telescope. The variation of the refraction index is larger (causing larger deviations of the beam) at shorter wavelengths. For this reason, the wavefront error (measured in units of the wavelength) is larger at shorter wavelengths.

These bubbles act as subpupils of aperture $r_0$ which form point spread functions (PSFs) with dimension $\theta_{atm} = 1.22 \frac{\lambda}{r_0}$. Having two bubbles at a distance among each other $\sim D$ constitutes a two-beam interferometer which creates a pattern of linear interference fringes \citep{1999PASP..111..587R}. Before the pupil telescope, there are sub-apertures of size $r_0$ and their fringes interfere constructively: these fringes constitute the variable ``speckle pattern''. The speckles have a size of $\sim \lambda/D$, for the distance of the two sources of interference. The speckles have the same dimension as the PSF of the star so it is impossible to distinguish them from a real source of the same brightness. 

To try to minimize this variable speckle noise, the AO uses a deformable mirror (DM) which permits to correct quasi-instantaneously the wavefront corrupted by the atmosphere. The wavefront sensor measures the wavefront corrugations, then a real-time control system calculates the correction to be sent to the DM, which has actuators that deform its shape. In order to model the atmospheric behavior, a point source, typically a star, is taken as a reference source. It has to be bright enough to permit a high signal-to-noise ratio (SNR) in the wavefront sensor. Some instruments have the possibility of using an artificial reference in addition to natural point sources, the laser guide star, when a bright star is lacking. Currently, most of the direct imaging instruments use the target host star itself as a reference for the AO. 

\begin{center}
	\begin{figure}[!htp]
	\centering
	\includegraphics[width=0.6\textwidth]{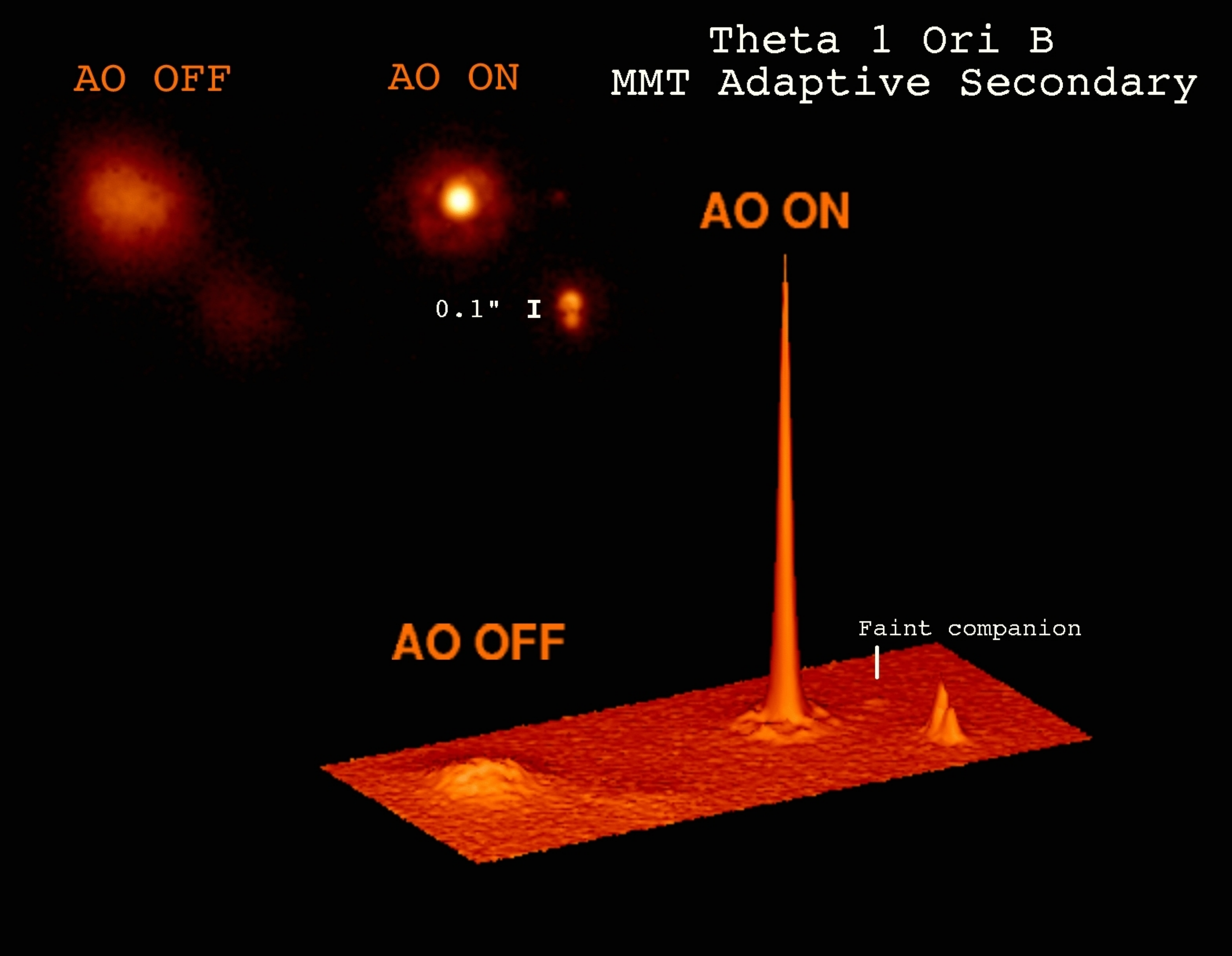}
	\caption{Differences between the performance with and without the adaptive optics. In the upper-left corner, the system of $\theta^1$ Orionis B is shown as seen in the detector of the instrument, with and without the AO. In the lower panel, the signal of the detector is shown in 3D, where the z-axis is the flux. Without the AO in the detector, two blurred blobs appear, when the AO is working, four objects are distinguishable. The SNR of each PSF increases when the AO is working. Credit: Laird Close, CAAO, Steward Observatory. }
        \label{f:AO}
	\end{figure}
	\end{center}

To commensurate the goodness of the AO, a parameter defined as Strehl Ratio (SR) is used. It defines the ratio between the peak of the PSF measured in the detector and the theoretical diffraction-limited PSF. In Fig.\ref{f:AO} the differences between the performance with and without the AO are shown.  A perfectly flat wavefront corresponds to the theoretical diffraction limit and then has SR$ = 100\%$. A corrugated wavefront has an SR $<$ 100\%. The larger the wavefront error, the lower the SR; for this reason, the Strehl ratio is higher at longer wavelengths. Also, as the Airy pattern dimension increases
as $\lambda$, and the image flux is being conserved, the higher SR (longer
wavelength) PSF has a brighter Airy pattern (core and inner rings) and a correspondingly lower intensity in its halo \citep{2000PASP..112...91M}.  While low-order AO can achieve SR of $20-60$\%, in the 1-2 $\mu$m wavelength range, extreme AO (ExAO, with a high number of actuators) can reach more than $90$\% SR, almost a diffraction-limited PSF, in the same wavelengths (see, e.g., the VLT/SPHERE manual\footnote{https://www.eso.org/sci/facilities/paranal/instruments/sphere/doc.html}). This improvement does not come for free, brighter stars are needed for the ExAO, as the wavefront must be more finely sampled. The ExAO provides a stabilized, quasi-diffraction limited beam, which is then centered in the coronagraph. To perform HCI the stability of the star PSF is key, to ensure that the host star is well centered and masked under the coronagraph during the whole exposure. 

The AO can correct the wavefront errors up to a separation defined as the outer working angle (OWA). It is the angular distance corresponding to spatial frequencies above $N_{act} \lambda/2D$ where $N_{act}$ is the number of actuators along a linear dimension of the DM and $D$ the telescope aperture. 

In addition to that due to the atmosphere, a second type of speckle pattern has to be taken into account: the quasi-static speckle pattern. The quasi-static speckles are due to the imperfections of the optics of the telescope, and they are mostly permanent during the exposure with typical lifetimes of tens of minutes \citep[see e.g.][]{2007ApJ...669..642S}. To attenuate this pattern, post-processing techniques are needed, as we will see in the next Section. The variable speckle pattern attenuated by the AO can be further suppressed during the post-processing by averaging the images during a long exposure. 

The speckles are brighter going closer to the host star and this is a limitation for the direct imaging, as the majority of the planets detected are close to their host (see Fig.~\ref{f:exo}). In fact, the major limitations of the direct imaging technique are the faintness of the companion with respect to the host star and its small angular separation. The angular separation is defined as the projected on-sky separation of the companion with respect to its host star.

\subsection{Coronagraphs}
HCI exploits different kinds of coronagraphs to attenuate the light of the primary, which are divided into many types \citep[see, e.g.,][]{2006ApJS..167...81G}. The main two categories are coronagraphs with occulting masks and phase masking coronagraphs. While the first class of coronagraphs uses an opaque mask that attenuates the first peak of the Airy function, the second class of coronagraphs exploits the principle of the destructive interference of the on-axis light. 

One of the principal parameters of the coronagraph is the inner working angle (IWA) which is the radius where the peak flux of the star is attenuated by 50\%. The goal of new coronagraphs is to decrease as much as possible this value, to be able to detect planets orbiting at small separations.  

\begin{center}
	\begin{figure}[!htp]
	\centering
	\includegraphics[width=\textwidth]{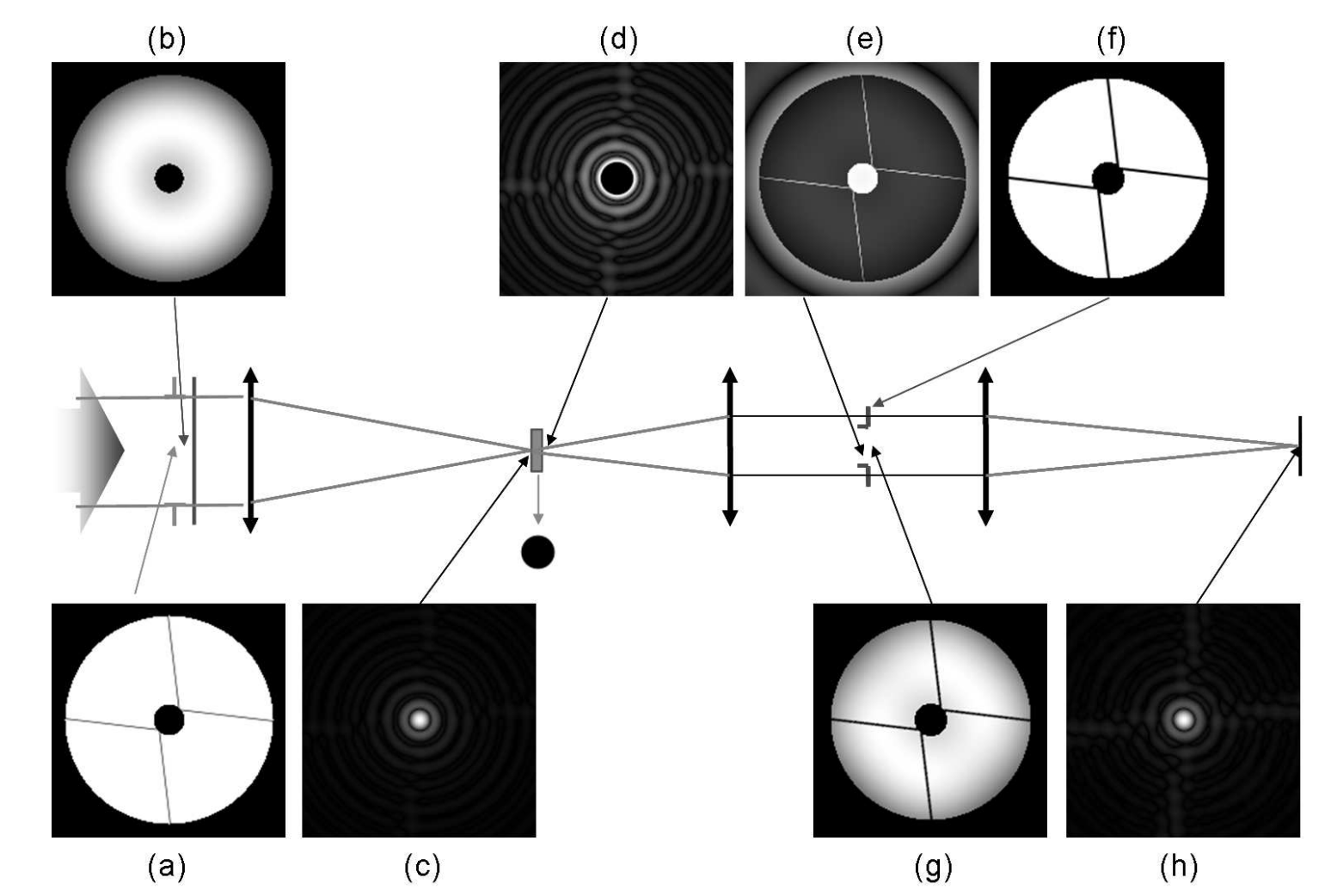}
	\caption{Principles of the apodized Lyot coronagraph (ALC) with a VLT pupil. Light enters from the left and passes through the various stages as shown. (a) Entrance pupil, (b) Apodizer, (c) Point spread function
(PSF) at the focus of the telescope, (d) PSF when the Lyot occulting coronagraphic mask is settled, (e) Pupil image
before the Lyot stop introduction, (f) Lyot stop, (g) Pupil image with the Lyot stop, (h) Final coronagraphic PSF Image. Image and caption from \citet{2009arXiv0901.2429G}. }
        \label{f:alc}
	\end{figure}
	\end{center}

The most common types of coronagraphs used by high-contrast imaging instruments are:
\begin{itemize}
\item Apodized Lyot coronagraph (ALC): It is part of the first generation of coronagraphs. It includes three optical elements: the occulting mask in the focal plane stops the light within a short separation from the center (typically $\sim$~0\farcs1 in 8-10 m telescopes); an apodizer located on a pupil plane (in front of the focal plane) limits the diffraction of the light with a gradual transmission function. A Lyot stop after the focal plane attenuates the diffracted light by the mask in the focal plane. It then has a shape similar to the occulting mask of the pupil plane, as shown in Fig.~\ref{f:alc} \citep[also see, e.g.,][]{2009arXiv0901.2429G}. 
\item Four quadrant phase-mask (4QPM): It is part of the second generation of coronagraphs. The mask (located in the focal plane) has a $\pi$ phase shift in each quadrant. When the beam of light is centered there is destructive interference that suppresses the light. This type of coronagraph is reasonably achromatic \citep[see, e.g.,][]{2000PASP..112.1479R}. 
\item Vector vortex coronagraph: It is part of the second generation of coronagraphs. The phase mask has different phases from 0 to 2$\pi$ which creates destructive interference of the light \citep[see, e.g.,][]{2009OExpr..17.1902M}. The vortex coronagraph is particularly efficient in providing a deep contrast at a small IWA. Currently, it is used for wavelengths longer than 3 \mic ~ (L band mostly). 
\item Apodizing phase plate (APP) coronagraph: It is part of the last generation of coronagraphs. The APP is a type of pupil-plane coronagraph that works by modifying the complex field of the incoming wavefront by adjusting only the phase \citep[see, e.g.,][]{2007ApJ...660..762K}. The advantage of this type of coronagraph is that the apodization is with phase only, which means that the throughput of the APP is higher compared to traditional
amplitude apodizers. Its improved version, the vector-APP, is currently installed in 6 high-contrast imagers \citep{2021ApOpt..60D..52D}. This version avoids chromatic biases and works in a wide wavelength range. 
\end{itemize} 

\section{Post-processing techniques}
\label{s:post}
To attenuate the quasi-static speckle pattern various techniques can be applied, as presented in this Section. The techniques presented here are applied to detect and characterize exoplanets, on the other hand, there is another technique used to detect circumstellar disks, the Polarized Differential Imaging (PDI). I refer the reader to \citet{2020A&A...633A..63D, 2020A&A...633A..64V} and references therein for a detailed description of this method, as it is not presented in this Chapter. 
\subsection{The SDI technique}
\label{s:sdi}
\begin{center}
	\begin{figure}[!htp]
	\centering
	\includegraphics[width=0.8\textwidth]{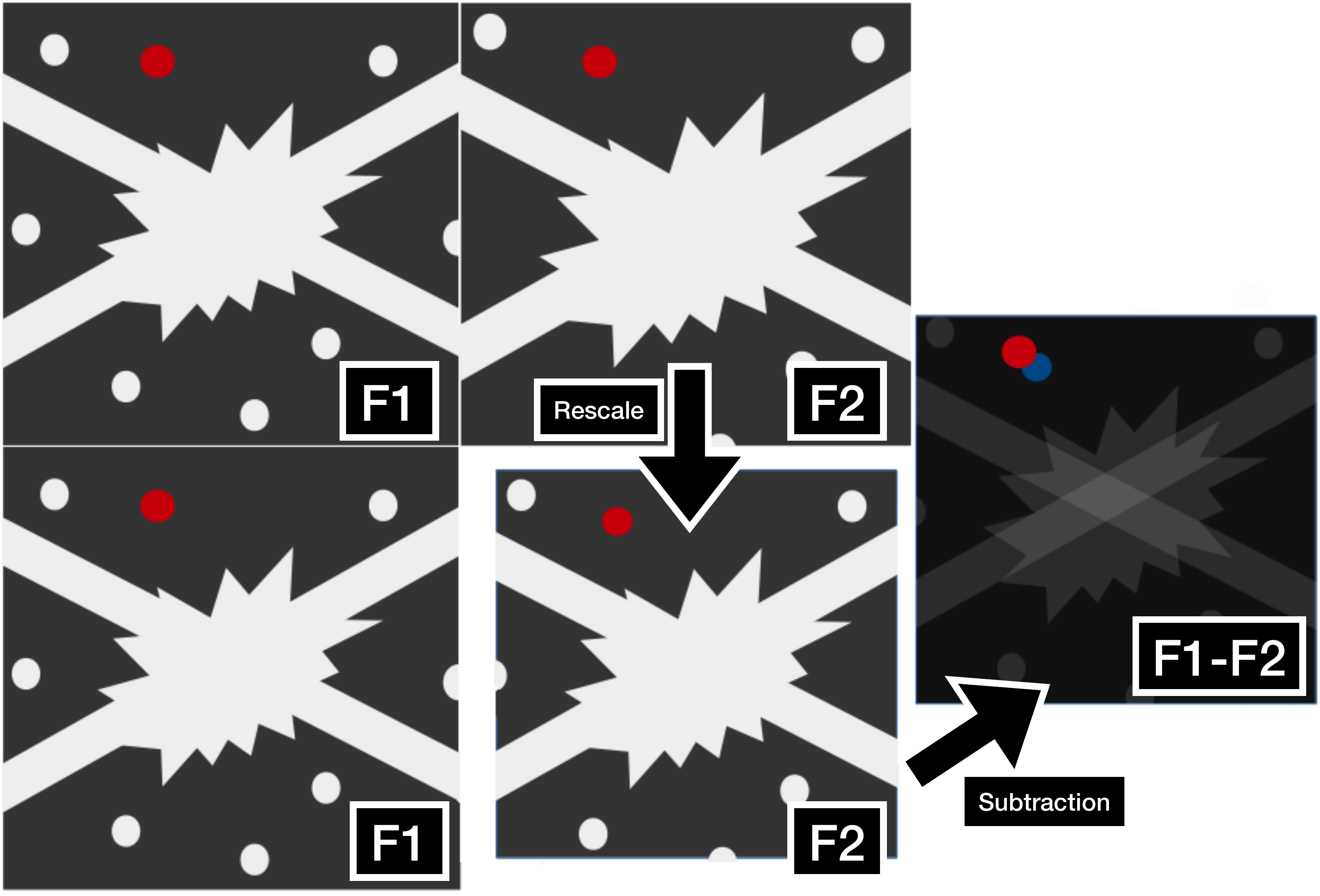}
	\caption{How to combine images to exploit the SDI technique. The top-left image represents the raw frame of filter1 (F1) and the top-right filter2 (F2). In the second raw filter1 is identical (lower left panel), while the image of filter2 has been rescaled to match the speckle pattern of filter1 (lower right panel). The product of the subtraction between the two filters after rescaling is shown on the right. The companion is represented in red, while its negative flux is represented in blue.}
        \label{f:sdi}
	\end{figure}
	\end{center}

This technique exploits the fact that the speckle pattern, which is a diffraction pattern (as we saw in Sec.~\ref{s:ins}), scales with the wavelength, while planets remain in the same position independently from the wavelength, as they are real objects in the observed field. The spectral differential imaging \citep[SDI,][]{1999PASP..111..587R} exploits this property to suppress the speckle pattern. The principle is to take simultaneously two coronagraphic images in two different bands with close wavelengths $\lambda_0$ and $\lambda_1$. In this way, the speckle pattern is almost identical in the two filters. The image is rescaled at the longer wavelength $\lambda_1$ to have the speckles at the same position in both images with respect to the center of the detector. On the other hand, the planet is shifted from the original coordinates, according to the wavelength ratio, and has a different position in the two images. The two images ($I$) are then subtracted according to the formula: 
\begin{equation}
I_{\rm diff}= I_{\lambda_0}- k(I_{\lambda_1})_{resc},
\end{equation}
where $k$ is a factor to correct for the flux amplitude of the speckle pattern, which can be different in the two filters. 
A cartoon of the SDI method is represented in Fig.~\ref{f:sdi}. 

\begin{center}
	\begin{figure}[!htp]
	\centering
	\includegraphics[width=\textwidth]{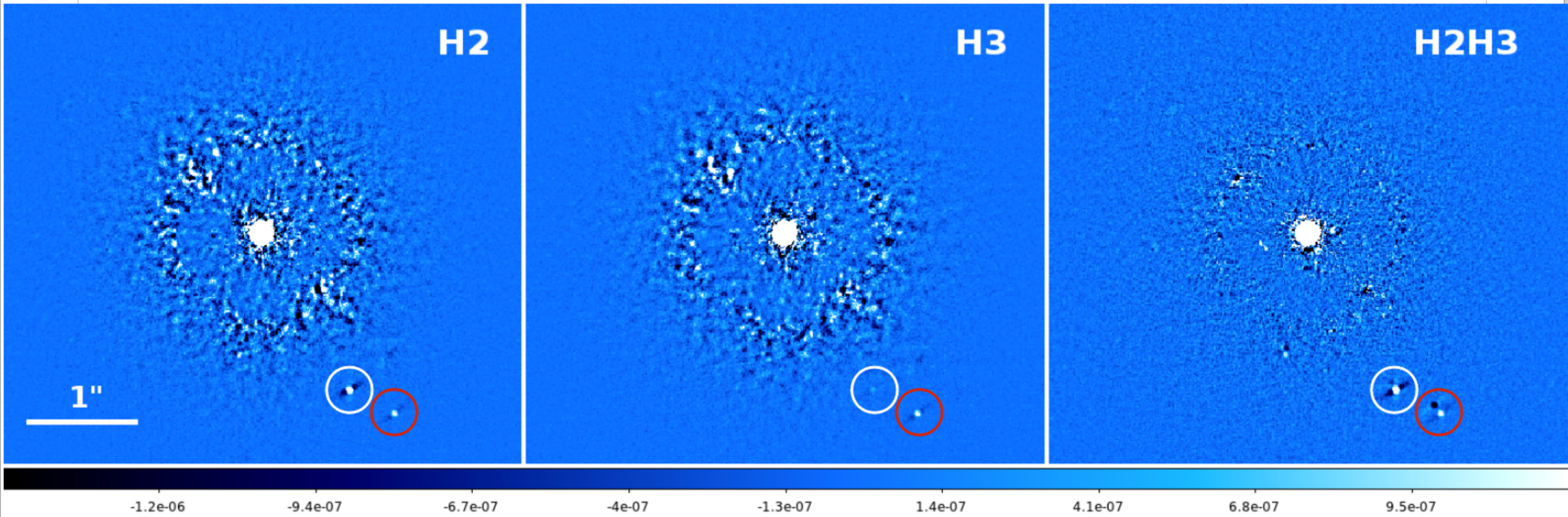}
	\caption{Images of the system around GJ\,758. The object in the white circle, GJ\,758\,B, is a late T-dwarf. In the red circle, a background star. The methane absorption in the $H3$ filter is evident, as the flux of the T-dwarf object is lower. In the SDI combination of the two filters, there is a clear difference between the T-dwarf and the background star close to it, the T-dwarf has a positive peak much brighter than the negative one, while for the star the two peaks are equally bright. Figure adapted from \cite{2016A&A...587A..55V}}
        \label{f:sdi_real}
	\end{figure}
	\end{center}

Also, this technique permits to distinguish a flat spectrum, where the fluxes are roughly the same in all the filters, from a spectrum that presents strong absorption in one of the channels, which will present different peak intensities in function of the wavelength. As we will see in Sec.~\ref{s:atm}, planetary spectra, especially T-type objects, present strong absorption in the bands of methane and water. For this reason, to observe in those bands, especially $J$ and $H$, is convenient to distinguish a planet from a background star with a flat spectrum, as shown in Fig.~\ref{f:sdi_real}. 

The filter pairs of planet-finder imagers are specifically selected to be on- and off- the molecular lines of interest for exoplanets, not only the methane lines (in the near-infrared) but also, for example, the H$\alpha$ line in the visible to detect planets that are actively accreting (as we will see in detail in Sec.~\ref{s:proto}).

\subsection{The ADI technique}
\label{s:adi}
Another fundamental observing method implied in HCI is the angular differential imaging \citep[ADI;][]{2006ApJ...641..556M}. This technique uses the intrinsic
field of view (FOV) rotation of altitude/azimuth telescopes to rotate the companions of the target around the center of the image, while the speckles remain as stable as possible during the observation (as we saw in Sec.~\ref{s:ins}). For an instrument at the Cassegrain focus\footnote{For an instrument at Nasmyth focus, the field rotates as the sum of the parallactic and altitude angle, while the pupil rotates as the altitude angle. A suitable optical element (e.g. a K-mirror) derotating the image as wished may then be located in front of the instrument, allowing stabilizing the image of the pupil on the detector; in this case, the field will rotate on the detector as the parallactic angle, as it is observed at the Cassegrain focus.}, the derotator of the telescope is switched off and the telescope pupil is fixed on the science camera. This observing setup is called ``pupil-stabilized'' mode. To exploit the maximum FOV rotation, the target is observed during the meridian passage, when the star is reaching its maximum altitude in the sky. The rotation rate $\psi$ (in degrees per minute) of the FOV is calculated as:
\begin{equation}
\psi = 0.2506~ \frac{\cos A \cos\phi}{\sin z},
\end{equation}
where $A$ is the target azimuth, $z$ the zenith distance and $\phi$ the telescope altitude \citep{1997eiad.conf.....M}.

\begin{center}
	\begin{figure}[!htp]
	\centering
	\includegraphics[width=\textwidth]{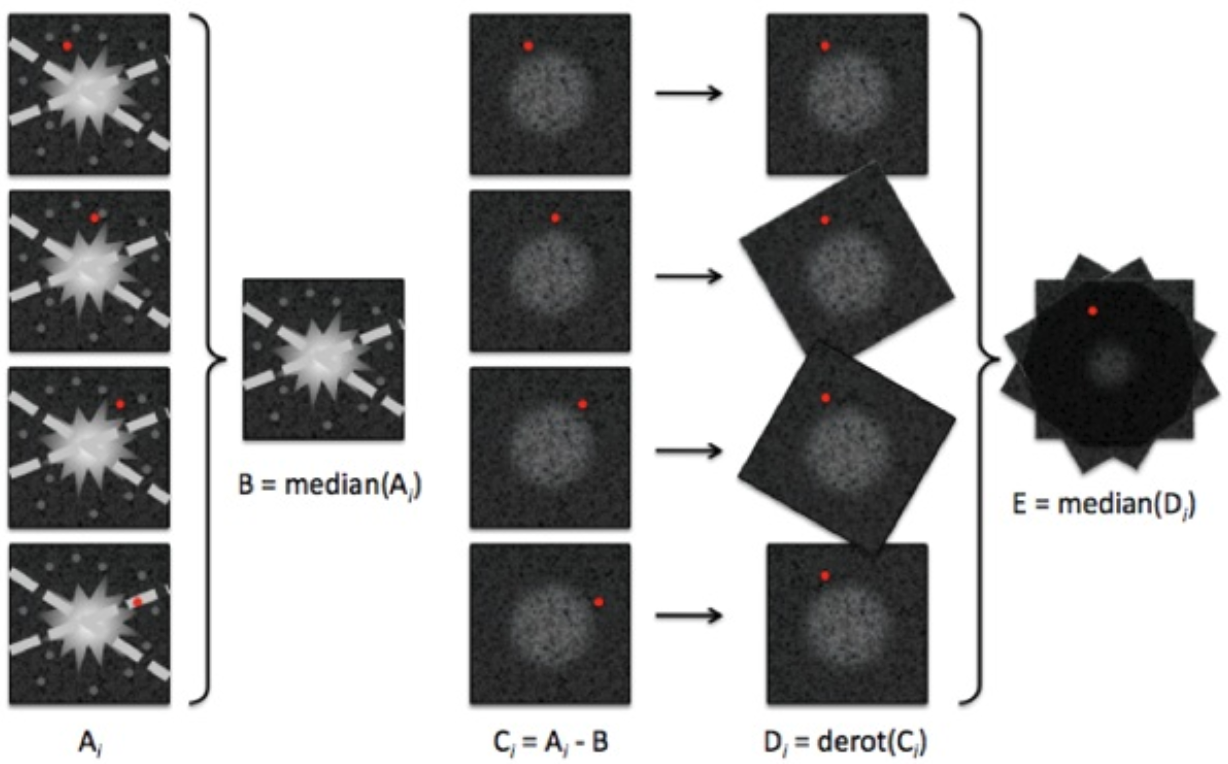}
	\caption{Images combination to exploit the ADI technique. In the first column, the raw frames are displayed (A$_i$). In the second column, the median of the raw frames is shown (B). The third column shows the same raw frames A$_i$ after the subtraction of the median of them (B). In the fourth column, each image has been derotated to have the North up and the East on the left (D$_i$). In the last column, the median E of the derotated frames (D$_i$) is displayed. The companion is represented in red. Credit: C. Thalmann.}
        \label{f:adi_ex}
	\end{figure}
	\end{center}

 In Fig.~\ref{f:adi_ex} the principle of a basic ADI is illustrated: the first step is to collect the images of the sequence, where the planet is slowly moving through the detector, while the speckles and the spiders are fixed (A$_{j}$); second, the images are combined with the median or other techniques to evaluate the speckles halo (B); then we subtract this product to the original images of the sequence (C$_{j}$); and derotate the images to have the planet at the same position (D$_{j}$) and the orientation with the North up and the East on the left; the result (E) is the combination of the derotated images.   

The efficiency of this technique improves for greater angular rotation of the FOV during the exposure. It is widely used by every direct imaging instrument. ADI can be combined with SDI to reach a deeper contrast. 

To post-process data from this observation technique there are many different reduction methods, among them I can cite:
\begin{itemize}
\item classical ADI \citep[cADI;][]{2006ApJ...641..556M}: as explained above;
\item smart ADI \citep[sADI;][]{2010Sci...329...57L}: a selection of the reference images which reconstructs the speckles halo is performed. While in the cADI all the images are taken into consideration and median-combined, in the sADI the closest-in-time frames to the reference image are excluded. This avoids the self-subtraction of the signal of the planet; 
\item radial ADI \citep[rADI;][]{2006ApJ...641..556M}: it is a procedure similar to the sADI. This time the selection of the frames is optimized for the separation of the object, as the signal of the planet rotates along the frame at different angles as a function of the radius planet-center of the image. The PSF of the planet moves faster along the images for larger separations (i.e. we can reject fewer frames). 
\item Locally Optimized Combination of Images \citep[LOCI;][]{2007ApJ...661.1208L}: it is an algorithm that constructs an optimized reference image to subtract from a set of reference images. This image is built as a linear combination
of the reference images selected, and the coefficients of the combination are optimized inside multiple subsections
of the image independently to minimize the residual noise within each subsection. There are also different flavors of LOCI, as for the ADI.
\end{itemize}

\subsection{The RDI technique}
\label{s:rdi}
The principle of Reference Differential Imaging (RDI) is that a model PSF of an isolated and disk-less reference star is taken and it is subtracted from the science frames. This technique is particularly useful for companions that are close to the host star \citep{2022A&A...666A..32X}. To fully take advantage of this method the "star-hopping" technique is being introduced, which consists of switching from the science target to the reference star multiple times during a sequence minimizing the observation overheads \citep[see, e.g.,][]{2021A&A...648A..26W}. Since a very stable PSF is required to perform the RDI, space telescopes are the ideal instruments to successfully apply this technique.     

\section{Planetary systems detected with direct imaging}
\label{s:di_systems}
The direct imaging technique has been providing us with discoveries of sub-stellar companions for twenty years, when the first planetary-mass companion, 2MASS J1207334-393254 (2M1207\,b, for simplicity) has been found with this method \citep{2005A&A...438L..25C}. Today around 30 objects in the planetary mass regime, M$_p \lesssim 13$ \MJup (see Sec.~\ref{s:atm}), have been discovered with this method, but the number will increase rapidly with the new generation of high-contrast imagers mounted in space or on extremely large telescopes. The planetary-mass companions discovered so far are listed in Table~\ref{t:di_sys}. All of them orbit far from their host stars, spanning semi-major axes from $10 \lesssim a \lesssim 6500$ au. One of the goals of the current (and future) generation of instruments is to explore regions as close as possible to the host star. This means that shorter-period objects than the ones discovered so far will be probably detected in the near future. Also, the achievable contrast is increasing, unveiling perhaps other additional companions around the targets that we previously did not have the capability to detect. Among the direct imaging discoveries, only three multiple planetary systems have been discovered so far, composed of four planets around the star HR\,8799 and two planets around both $\beta$ Pictoris and PDS\,70. 

The results of HCI campaigns yield statistical results, even in the case of non-detection. For example, \citet{2007ApJS..173..143B} and \citet{2008ApJ...674..466N} found with 95\% confidence level that less than 20\% of stars have planets with a mass greater than 4 \MJup between 20 and 100 au. \citet{2007ApJ...670.1367L} estimated that less than 17\% of stars have planets with a mass between 0.5 and 13 \MJup on semi-major axes between 25 and 325 au, and at most 10\% between 50 and 220 au. \citet{2015A&A...573A.127C} constrained the occurrence of planets more
massive than 5 \MJup to less than 15\% between 100 and 300 au, and for companions more massive than 10 \MJup to less than 10\% between 50 and 300 au. 
From GPIES, \cite{2019AJ....158...13N} found a clear correlation between the mass of the host star and the planet occurrence rate,  with stars $>$ 1.5 \MSun more likely to host giant planets (5–13 \MJup) at wide separations (semimajor axes 10–100 au) than lower-mass stars. Around higher-mass stars, the total occurrence rate of such planets is 9\%. \cite{2021A&A...651A..72V} found that the frequencies of systems with at least one substellar companion with masses between 1 and 75 \MJup and semimajor axes between 5 and 300 au are 23.0, 5.8, and 12.6\% for BA, FGK, and M stars, respectively. The surveys mentioned above targeted hundreds of young and nearby stars, and among the planets discovered, some benchmark objects deserved particular attention and dedicated observations. In the following sections, I will present some of the well-studied systems discovered so far.

\begin{center}  
\begin{table}
\caption{Planetary confirmed companions (M $\lesssim$ 13 \MJup) detected by the direct imaging technique. The distances to the systems are taken from Gaia collaboration et al. (2020).} 
\label{t:di_sys}
\vspace{0.2mm}
\begin{center} 
\renewcommand{\footnoterule}{}  
\begin{tabular}{llllll}
\hline
\hline
Planet   &  M (\MJup) & a (au) & dist (pc) & Discovery paper \\
 \hline
2M1207\,b & 5  & 42 & 65 & {\cite{2005A&A...438L..25C}}\\            
AB\,Pic\,b & 10--14   & 260 & 50 & \cite{chav2005}  \\
DH\,Tau\,b & 11 & 330 &  134 & \cite{2005ApJ...620..984I} \\
HR\,8799\,b  & $\sim 5$  & 69.2 &  41 & \citet{2008Sci...322.1348M} \\
HR\,8799\,c  & $\sim 7$ & 37.4  & 41 &\citet{2008Sci...322.1348M} \\
HR\,8799\,d  &$\sim 7$ & 24.5 & 41 & \citet{2008Sci...322.1348M} \\
1RXS\,1609\,b & 8-14 & 330 & 145 &\citet{2008ApJ...689L.153L}\\
$\beta$\,Pic\,b   & 9  & 10  & 19\footnote{from \cite{2007A&A...474..653V}} &\citet{2010Sci...329...57L}   \\
HR\,8799\,e  &$\sim 7$& 14.5 & 41 & \citet{2010Natur.468.1080M} \\
2MJ0441+23\,b & 7.5 & 15 & 140  &\citet{2010ApJ...714L..84T}  \\
Ross\,458\,c & 11 & 1168 & 11.7 &\citet{2010ApJ...725.1405B}   \\
WD\,0806-661\,b & 8 &  2500 & 19.2 & \citet{2011ApJ...730L...9L}     \\
HD\,95086\,b &    $\sim 3$     & 56 &  86   & \citet{2013ApJ...772L..15R} \\
Gliese\,504\,b & 4 & 43.5 &  18  &\citet{2013ApJ...774...11K} \\
$\kappa$\,Andromedae\,b & 13 & 100  & 50  & \cite{2013ApJ...763L..32C} \\
2MASS\,J0103\,AB\,b & 13 &  84 & 47   &{\cite{2013A&A...553L...5D}}\\
ROXs\,42\,(AB)b & 9 & 140 &  135  &\citet{2014ApJ...780L..30C} \\
HD\,106906\,(AB)b& 11 & 850	&  92 & \citet{2014ApJ...780L...4B}  \\
GU\,Piscium\,b& 11&	2000 &	48&\citet{2014ApJ...787....5N}  \\
51\,Eri\,b & 3 & 11  & 29 & \cite{2015Sci...350...64M}\\ 
HIP\,65426\,b & 7 & 87 & 109 & {\cite{2019A&A...623A.140G}} \\
2MASS\,J22362452+4751425 & 12 & 230 & 63 & {\cite{2017AJ....153...18B}} \\
PDS\,70\,b  & 7  &  22 &  113  &\cite{keppler}\\
PDS\,70\,c &  4 &  30 &  113  &  \cite{2019NatAs...3..749H} \\
HD\,169142\,b & 2 & 36 &117 & {\cite{2019A&A...623A.140G,2023MNRAS.522L..51H}} \\
2M0437\,b & 4 & 118 & 128& \cite{2022MNRAS.512..583G} \\
YSES\,2\,b & 6 & 114 & 109 & {\cite{2021A&A...648A..73B}} \\
COCONUTS-2\,b  & 6.3 & 6471 & 11 &  \cite{2021ApJ...916L..11Z} \\
AB\,Aurigae\,b & 9--12 & 94 & 144 & \cite{2022NatAs...6..751C} \\
AF\,Lep\,b&2--5.5 &20.6 &8&{\citet{mesa,2023A&A...672A..94D,2023ApJ...950L..19F}}\\
HIP\,99770\,b & 13-16 & 16.9 & 41 & \cite{2023Sci...380..198C} \\
\hline
\end{tabular}
\end{center}
\end{table}
\end{center}

\subsection{2M1207}
\begin{center}
	\begin{figure}[!htp]
	\centering
	\includegraphics[width=8cm]{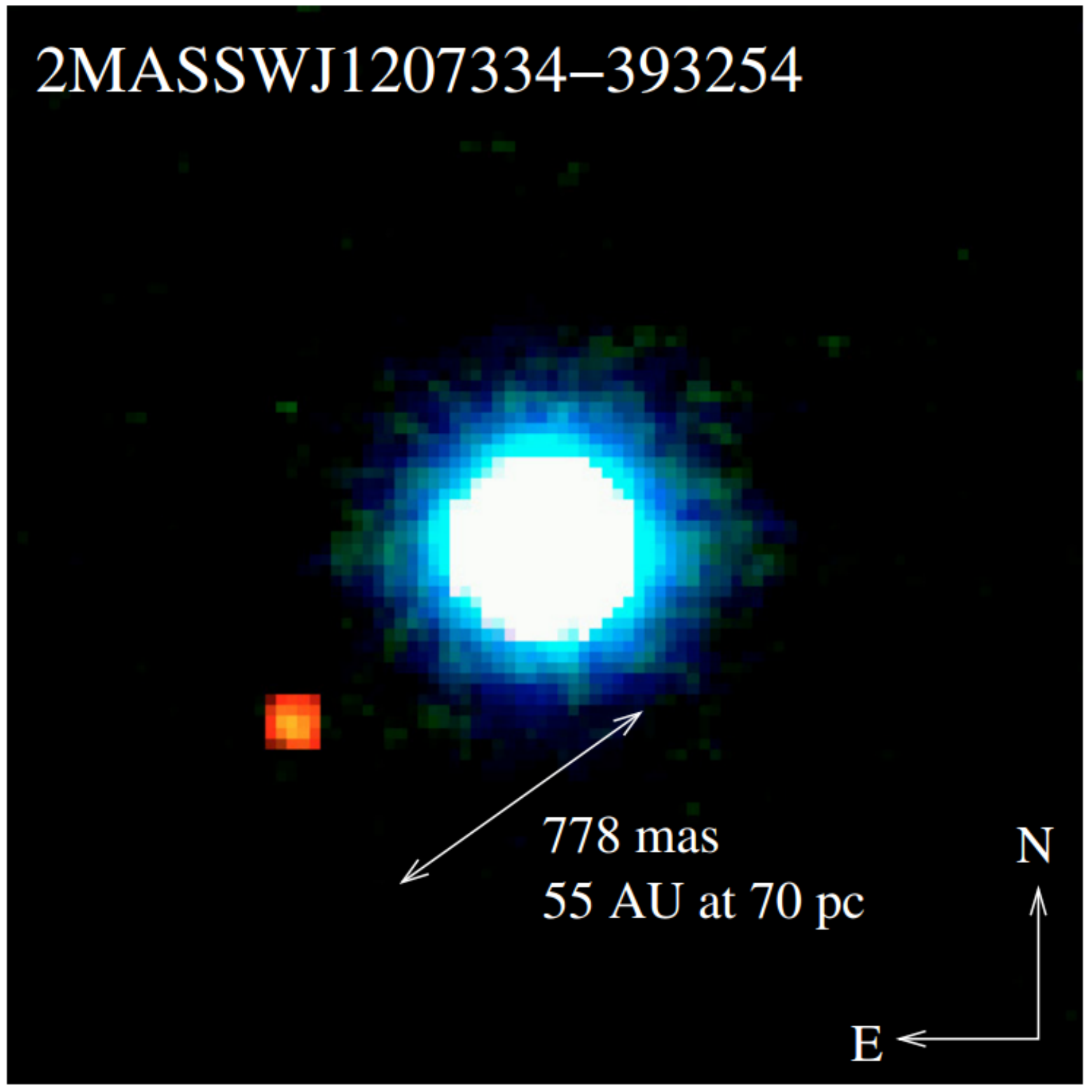}
	\caption{The first planet ever detected with the high-contrast imaging technique: 2M1207\,b \citep{2004A&A...425L..29C}.}
        \label{f:chauv}
	\end{figure}
	\end{center}

2M1207\,b is the first exoplanet detected with the direct imaging technique, it has been discovered using the instrument NACO, installed at the VLT \citep{2005A&A...438L..25C}. The first epoch was obtained in 2004 when a promising object was detected around the star 2MASSW J1207334-393254 (Fig.~\ref{f:chauv}), an M8 type star of the TW Hydra association (age $\sim 8$ Myr). The object is separated from the host star by about 0\farcs78 (= 55 au). To confirm that the object is comoving, a second epoch was taken one year later, and the nature of the planet was confirmed. Following evolutionary models, a mass of M $= 5\pm 2$ \MJup and an effective temperature of $T_{eff} = 1250 \pm 200~K$ is found. It is considered more a binary system of low-mass objects, rather than a planetary system \citep{2007ApJ...657.1064M}.  

\subsection{HR\,8799}
\begin{center}
	\begin{figure}[!htp]
	\centering
	\includegraphics[width=12cm]{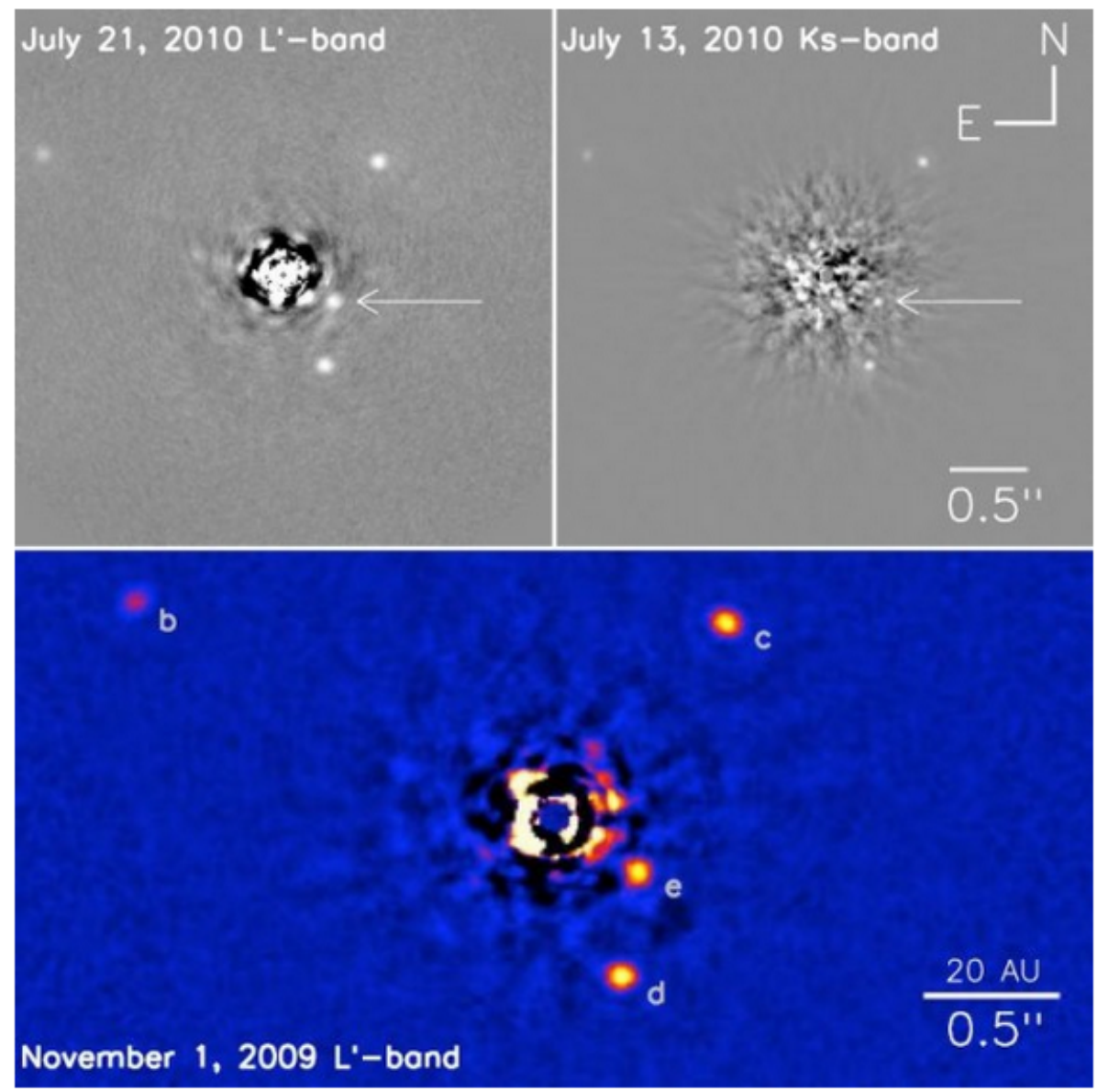}
	\caption{The first multi-planetary system ever imaged: HR\,8799\,bcde \citep[from][]{2010Natur.468.1080M}.}
        \label{f:hr8799_det}
	\end{figure}
	\end{center}

High-contrast observations with the Keck and Gemini telescopes have revealed in 2008 the first multi-planetary system ever imaged: four planets
orbiting the star HR\,8799 \citep{2008Sci...322.1348M,2010Natur.468.1080M}. One of the first images of the system is shown in Fig.~\ref{f:hr8799_det}. 
HR\,8799 is a young star ($\sim$ 42 Myr), $\gamma$ Doradus variable with $\lambda$ Boo-like abundance patterns. The distance to the system is 40 pc \citep{2020yCat.1350....0G}.

This system is the most observed target with the direct imaging technique and a benchmark for the study of planetary atmospheres, dynamics of multi-planetary systems, and formation mechanisms. HR\,8799 is particularly suited for this kind of observation since the contrast between the planets and the central star is favorable. The planets are also orbiting on a wide separation from the host star, making direct imaging observation relatively easy for most of the current high-contrast instruments. 

For this reason, the planets around HR 8799 have been observed by different telescopes and instruments, providing a rich archive of astrometrical, photometrical, and spectroscopical information, invaluable to studying young planetary systems. 

Astrometric follow-ups of the planets lasted two decades, permitting monitoring of the orbits of the 4 planets. In \cite{2022A&A...666A.133Z} a detailed dynamical and orbital analysis is presented. The parameters for the 4 orbits are calculated assuming coplanarity, relatively small eccentricities, and periods very close to the 2:1 resonance. The dynamical masses for the planets have been estimated to be 8--9 \MJup for the inner planets HR\,8799edc, and 6 \MJup for planet b. We refer the interested reader to these publications on the astrometrical characterization of the system:  \citet{2012ApJ...755...38S, 2012ApJ...755L..34C,2013A&A...549A..52E,2015A&A...576A.133M,2015ApJ...803...31P, 2016A&A...587A..57Z,2016AJ....152...28K,2018AJ....156..192W,2019A&A...623L..11G}.

Spectroscopy of the planets has been performed extensively and covers a wide wavelength range. The spectra recovered measure the emission of the planets themselves, without contamination from the host star. Among the results presented, I can cite \citet{2012ApJ...753...14S, 2012ApJ...754..135M, 2013ApJ...768...24O,2014ApJ...794L..15I,2018AJ....155..226G,2020MNRAS.491.1795P,2021AJ....162..290R,2022AJ....164..143W}. The variability of the atmospheres has also been monitored, although no evident signal of variability has been found \citep{2016ApJ...820...40A,2021MNRAS.503..743B}

\subsection{Fomalhaut}
\begin{center}
	\begin{figure}[!htp]
	\centering
	\includegraphics[width=0.7\textwidth]{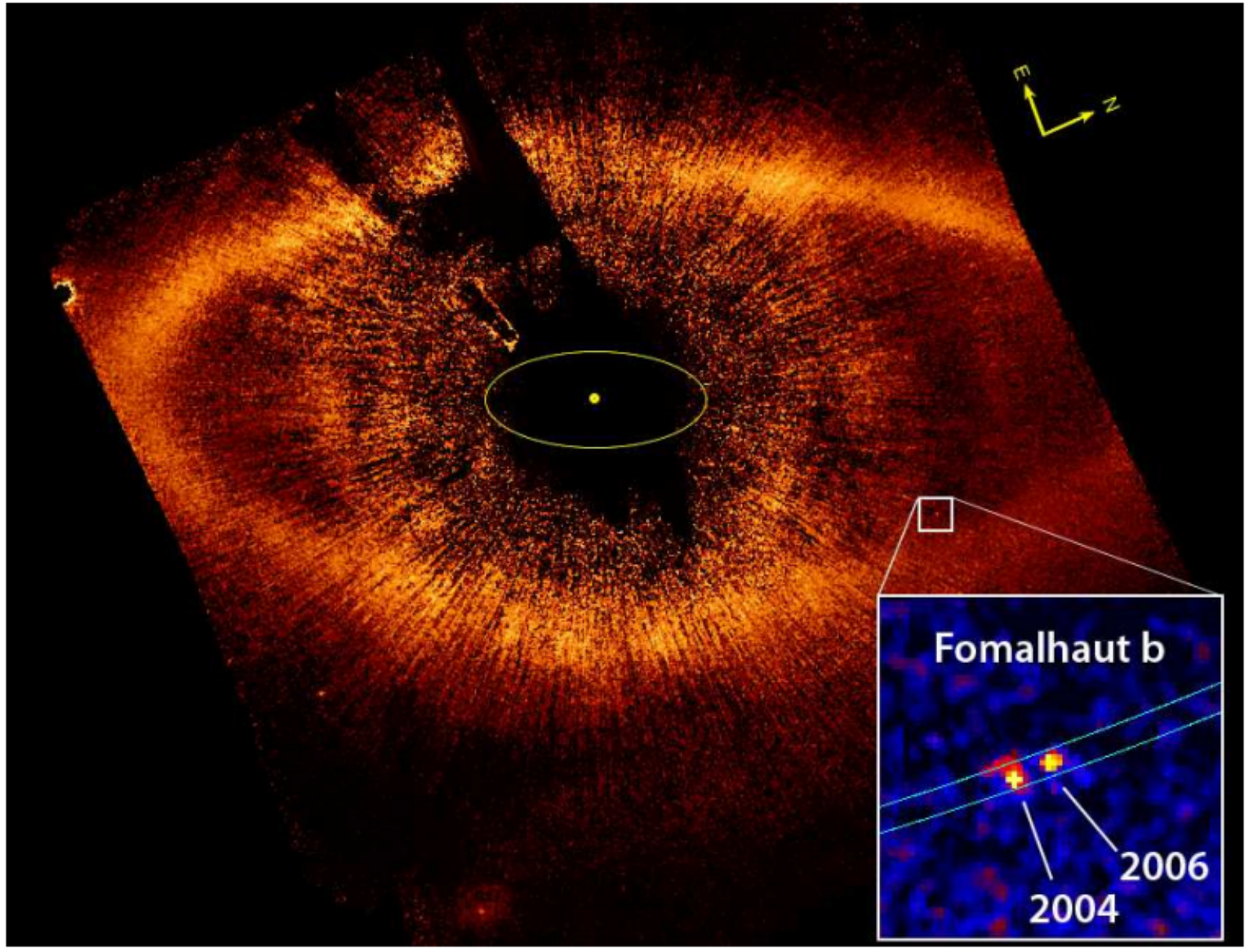}
	\caption{Image of the debris disk surrounding the star Fomalhaut. The putative planet Fomalhaut\,b is seen in two different epochs \citep{2008Sci...322.1345K}.}
        \label{f:foma}
	\end{figure}
	\end{center}

\citet{2008Sci...322.1345K} discovered the lowest mass planet ever imaged, Fomalhaut\,b (Fig.~\ref{f:foma}). This object has been imaged in visible light using Hubble Space Telescope (HST) Advanced Camera for Surveys (ACS), and it appears in the dust belt surrounding the brightest star of the constellation Piscis Austrinus. Fomalhaut is an A4 type star, 7.7 pc away, aged $\sim 400$ Myr. The planet is not detected in IR light. Neither Keck ($L^{\prime}$ band) nor Gemini ($H$ band) have been able to detect it. This can suggest that the planet is (a) low mass and (b) detected in the visible through reflected light more than thermal emission. Its presence was predicted by the unusual sculpturing of the disk, which is highly inclined and off-centered. The origin of the planet is accepted to be in the debris disk, as it is still surrounded by it.
Dynamical studies put an upper limit for the mass of the object of 3 \MJup. Fomalhaut\,b is on a very eccentric (0.8) orbit, with a semi-major axis of 177 au \citep{2013ApJ...775...56K}. It should be mentioned that the nature of this object is still under debate, with the alternative that the object is instead a dust clump \citep[see, e.g.,][]{2013ApJ...769...42G, 2013ApJ...775...56K}. Lately \citet{2015MNRAS.448..376N} proposed that the companion is, in reality, a background neutron star, confirmed by \citet{2023MNRAS.524.2698K}. 

\subsection{$\beta$\,Pictoris}
\begin{center}
	\begin{figure}[!htp]
	\centering
	\includegraphics[width=12cm]{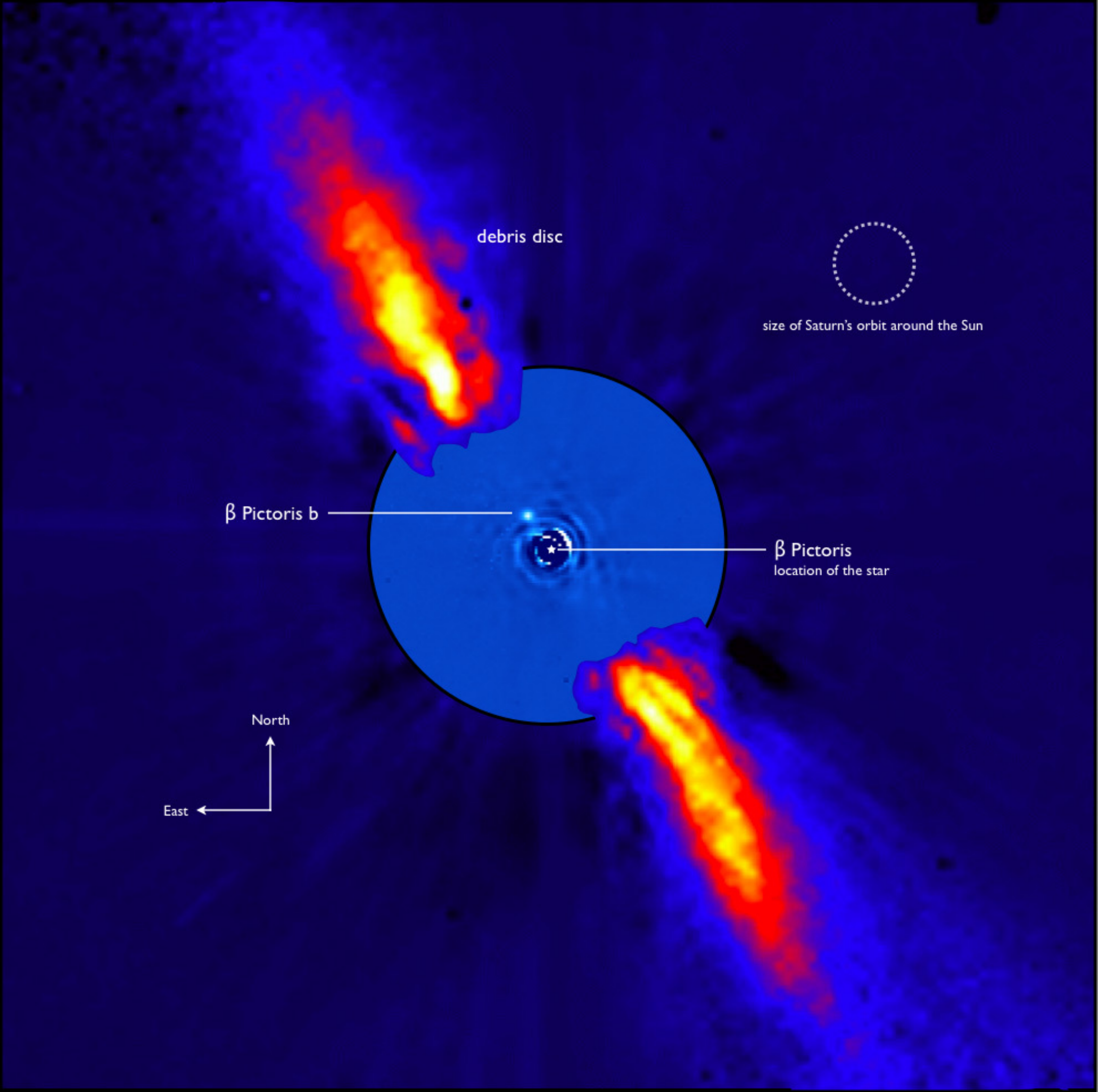}
	\caption{Image of $\beta$\,Pic, showing planet b and the debris disk around it \citep{2009A&A...493L..21L}.}
        \label{f:beta_lag}
	\end{figure}
	\end{center}

Around the star $\beta$\,Pictoris, a planet has been discovered by \citet{2009A&A...493L..21L}. The system also hosts an edge-on debris disk, as shown in Fig.~\ref{f:beta_lag}. This planet has the closest semi-major axis for a direct imaging planet, $8-10$ au, that corresponds to the snow line of the system, as the primary is an A-type star. As for the Fomalhaut planetary system, upper limits on the mass of the companion can be calculated from dynamical constraints. The mass of $\beta$\,Pic\,b has been estimated of $\sim$ 8 \MJup. The age of the system is assumed to be 18 Myr, as in \citet{2020A&A...642A.179M}. The small projected separation of the system, 0\farcs4, is no longer an issue for the new generation instruments, which are finally able to extract the spectrum of the companion \citep[see, e.g.,][]{2020A&A...633A.110G}.

$\beta$ Pic turned out to be a multi-planetary system, with the radial velocity technique, a second companion was discovered \citep{2019NatAs...3.1135L}. It orbits closer to the star, with a semi-major axis of 2.7 au. The second companion to the system, $\beta$ Pic\,c, was directly confirmed using the VLTI/GRAVITY instrument and presented in \cite{2020A&A...642L...2N}. $\beta$ Pic\,c is the first planet discovered with the radial velocity technique and confirmed by direct detection. Having both information on the radial velocity and luminosity of the planet can constrain the dynamical mass of the object, and it is invaluable information to confirm the predictions of evolutionary models. Its mass is about 8~\MJup \citep{2020A&A...642A..18L}. 

\subsection{PDS 70}

\begin{center}
	\begin{figure}[!htp]
	\centering
	\includegraphics[width=11cm]{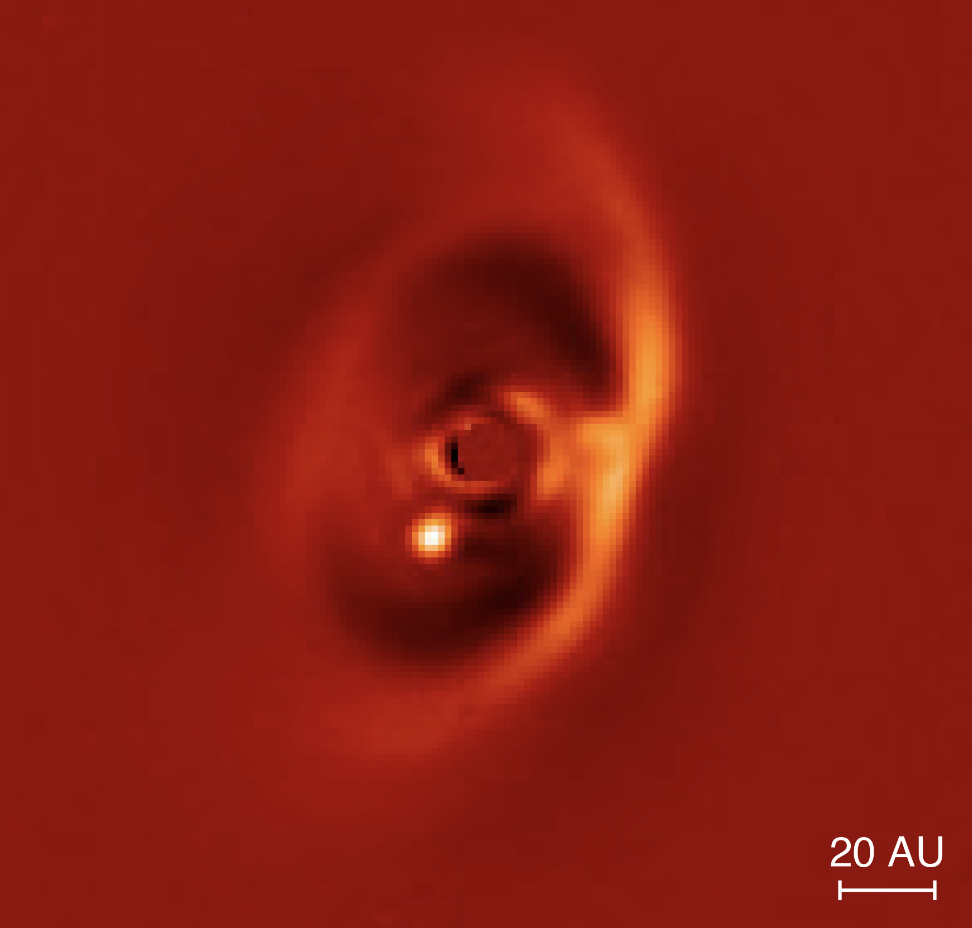}
	\caption{Image of PDS\,70, showing the first detection of planet b and the cavity of its protoplanetary disk \citep{keppler}.}
        \label{f:kepp}
	\end{figure}
	\end{center}

PDS 70 is one of the most emblematic systems discovered with the direct imaging technique. The host star is a K7-type member part of the Scorpius-Centaurus association \citep{2016MNRAS.461..794P}. The distance to the system is 113.47 pc \citep{2020yCat.1350....0G}. The star is surrounded by a protoplanetary disk with a wide cavity. For the first time, a planet was discovered inside the gap of the disk, PDS 70 b \citep[Fig.~\ref{f:kepp} and][]{keppler}. For the discovery, the instruments SPHERE and NACO at the VLT were used. Given the young age of the system, 5.4 $\pm$ 1.0 Myr, the mass of the planet was estimated around 5--9 \MJup \citep{keppler,2018A&A...617L...2M}. Planet b was then imaged in the H$\alpha$ filter, demonstrating signs of accretion \citep[][see also Sec.~\ref{s:proto} on the implication of this discovery]{2018ApJ...863L...8W}. A subsequent follow-up with the instrument MUSE revealed that the star hosts two planets, the other planet c was detected at a separation of 30 au, and both planets are actively accreting. This discovery was crucial as it provided direct observational evidence of ongoing planet formation in a protoplanetary disk. The planets that are still in the process of accreting material from the surrounding disk are called ``protoplanets'' (as we will see in Sec.~\ref{s:proto}).

\begin{center}
	\begin{figure}[!htp]
	\centering
	\includegraphics[width=11cm]{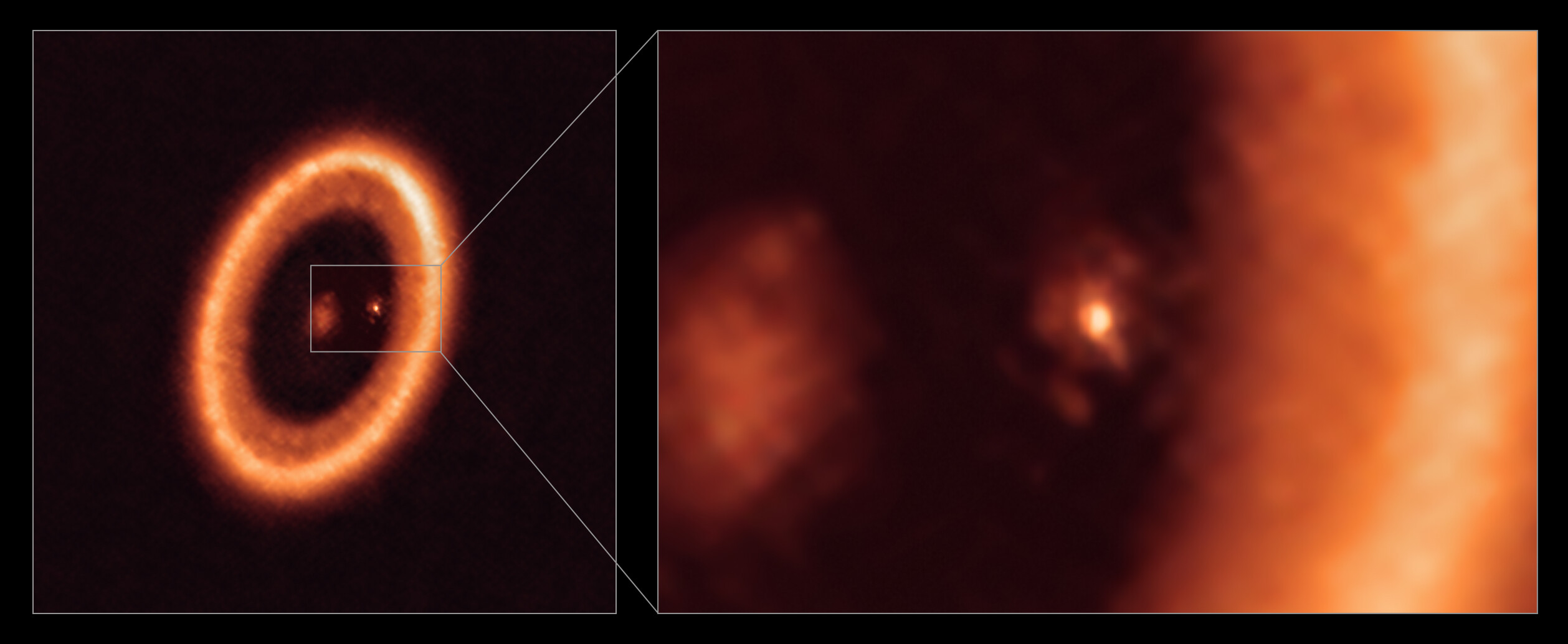}
	\caption{Image of the circumplanetary disk around PDS\,70\c, the first ever detected \citep{2021ApJ...916L...2B}.}
        \label{f:ben}
	\end{figure}
	\end{center}

The system was also imaged at longer wavelengths to study the disk properties \citep{2019ApJ...879L..25I}. In 2021, a groundbreaking discovery was announced, adding even more focus to this emblematic system. At very high resolution, for the first time, a circumplanetary disk was detected around planet c \citep[Fig.~\ref{f:ben} and][]{2021ApJ...916L...2B}. Future follow-ups on the circumplanetary disk detected around planet c, or new detections around other companions, will provide information on the formation of satellites around giant planets.

\subsection{AF Lep}

\begin{center}
	\begin{figure}[!htp]
	\centering
	\includegraphics[width=11cm]{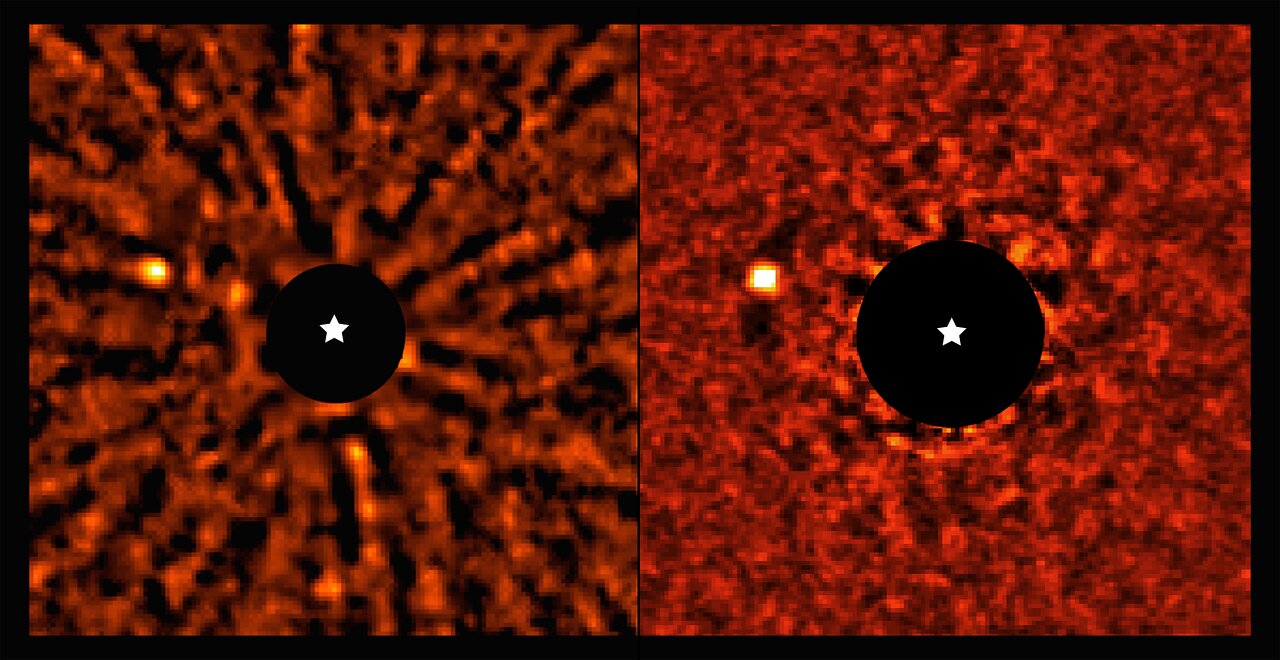}
	\caption{The VLT/SPHERE detection of AF\,Lep\,b \citep[Credit: ESO and][]{mesa,2023A&A...672A..94D}.}
        \label{f:aflep}
	\end{figure}
	\end{center}

AF Leporis is a star of the $\beta$ Pic moving group. Very recently, three independent publications, \citet{mesa,2023A&A...672A..94D,2023ApJ...950L..19F} discovered a Jupiter-like planet around the star (Fig.~\ref{f:aflep}). This is the third discovery in the same moving group after $\beta$ Pic\,b and 51\,Eri\,b. The age of the system is the same as the other stars in the moving group, around 20 Myr. The star was observed with high-contrast imaging for being an accelerating star. The presence of a companion was foreseen by comparing the astrometric values taken by Hipparcos and Gaia. The mass of the planet is estimated to be smaller than 5 \MJup, on an orbit of a= 21 au. It is one of the lowest-mass planets detected with the direct imaging technique. Once again, combining the results from two different techniques, HCI and the Gaia-Hipparcos acceleration, provided stronger constraints on the mass determination of the companion.  

\section{Mechanisms of giant planet formation}
\label{s:form}

Despite the wealth of exoplanets detected, numerous questions regarding planet formation remain unanswered, such as the processes and locations where planets originate. The two most widely accepted scenarios for planet formation are the \textit{core accretion} \citep[CA,][]{1980PThPh..64..544M, 1996Icar..124...62P} and the \textit{gravitational instability} \citep[GI,][]{1978M&P....18....5C, 1997Sci...276.1836B} models.

\subsection{Core accretion model}
One of the theories of planet formation most widely accepted is the so-called {\it core accretion} model \citep{1980PThPh..64..544M, 1996Icar..124...62P}. It is the mechanism mostly accepted for the formation of giant gas planets in our solar system and in exoplanetary systems, at least for objects orbiting closer than 10-50~au \citep{2014exha.book.....P}. This scenario is based on the fact that in the protoplanetary disk solid objects, called {\it planetesimals}, are present. These objects can collapse on a solid core, created from dust if located inside a zone where the protoplanet has gravitational influence. The solid cores can be formed more easily in the outer part of the disk, beyond the snow line, where the temperature is low enough for the formation of water ices. The feeding zone can extend over a few Hill radii $R_{H} = a(M_{p}/ 3 M_{\star})^{1/3}$, where $a$ is the planet orbital radius, $M_{p}$ and $M_{\star}$ is the protoplanet and star mass respectively \citep{2010RPPh...73a6901B}. When the solid core (ice and rock) has reached a mass of $\sim 0.1$~M$_{\oplus}$ it starts to gather an envelope of nebular gas. In this phase, there is a quasi-static balance between radiative loss and accretion energy. When the gas accretion has reached a critical mass $M_{crit}$ a gravitational contraction occurs to compensate the radiative loss. The critical mass is reached when the mass of the core and the envelope are roughly equal \citep{2008ASPC..398..235M}. The critical mass can be estimated to be in the range 5-15 M$_{\oplus}$ depending
on physical conditions and assumptions about grain opacity \citep[see, e.g,][]{1996Icar..124...62P}. When the critical mass is reached, the envelope can stay no longer in hydrostatic equilibrium. It begins to contract and the gas falls in free fall onto the core. The radius of the newborn planet is fixed by the conditions of this radiative shock. The accretion process terminates when the planetesimals and gas supplies end, as a gap in the disk may be created or because the disk gas dissipates at some point. 

\begin{center}
	\begin{figure}[!htp]
	\centering
	\includegraphics[width=0.45\textwidth]{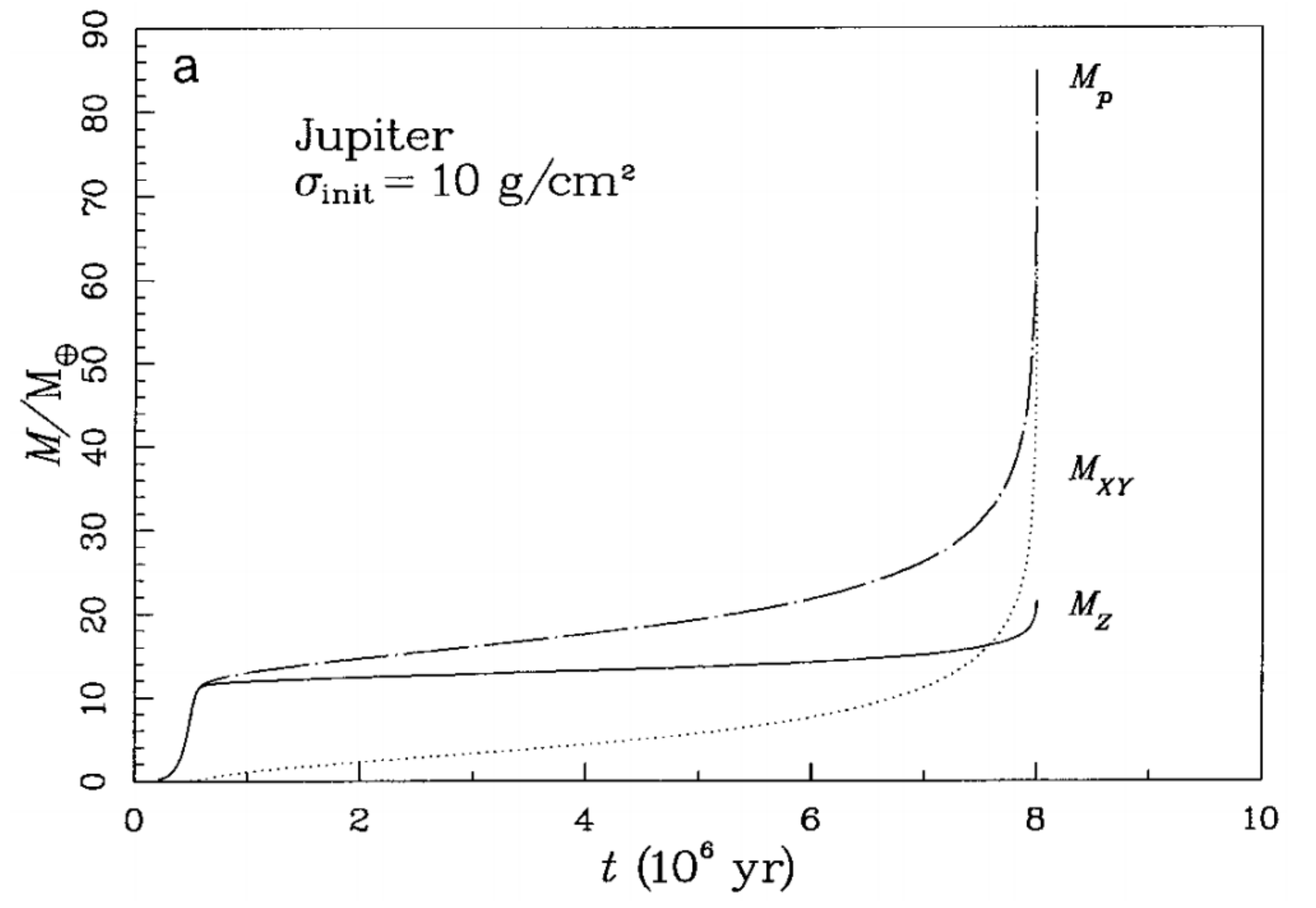}
\includegraphics[width=0.45\textwidth]{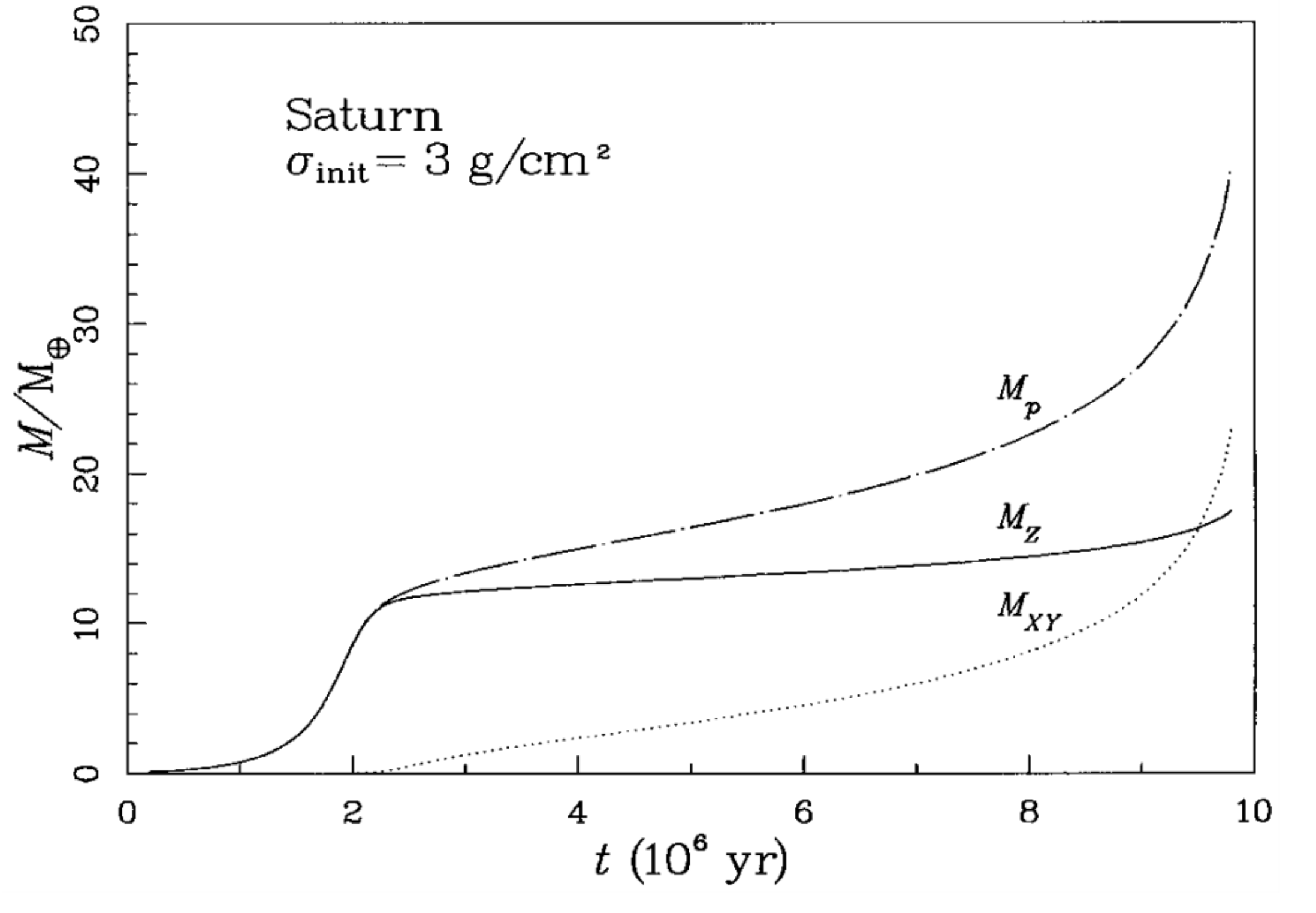}
\includegraphics[width=0.45\textwidth]{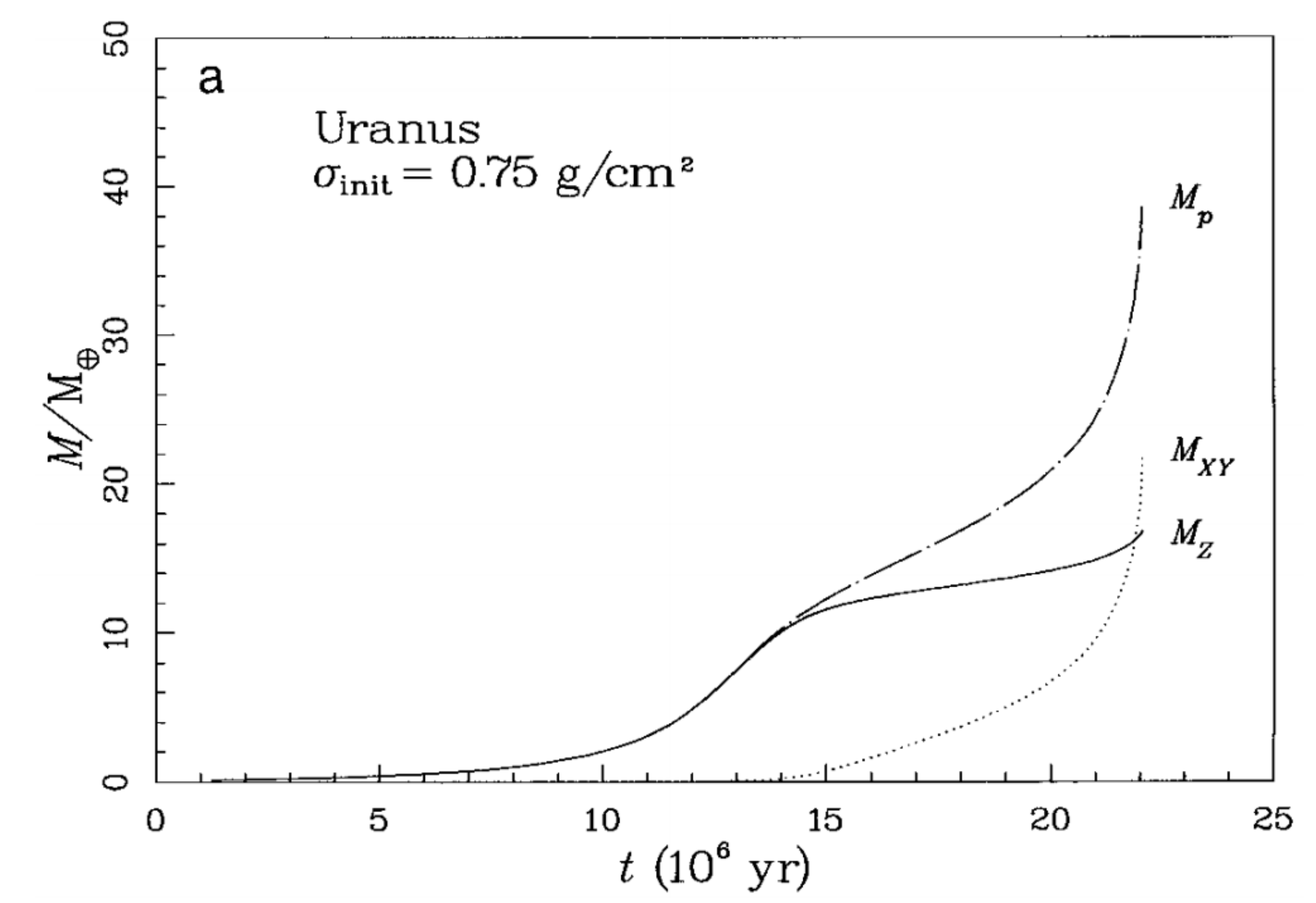}
	\caption{Figures 1, 4, and 5 from \citet{1996Icar..124...62P}. Results on the simulations performed on the formation of the solar system giant planets. The initial planetesimal surface density, $\sigma_{init}$ is shown for each planet. The initial embryo has nearly the mass of Mars, and planetesimals have a radius of 100 km. The solid line (M$_Z$) represents accumulated solid mass, the dotted line (M$_{XY}$) accumulated gas mass, and the dot-dashed line (M$_P$) the planet’s total mass.}
\label{f:pollack}	
\end{figure}
	\end{center}

An example of a simulation of Solar system giant planets formation is given by \citet{1996Icar..124...62P}. The process is divided into three phases: (i) the embryo of the protoplanet accumulates planetesimals in the first 5$\times$10$^{5}$~yr, and there is a depletion of its feeding zone; (ii) the accretion rate remains constant during $\sim$ 7~Myr, and the growth accelerates during the beginning of the phase (iii) with a runaway accumulation of gas. The results of these simulations are shown in Fig.~\ref{f:pollack}. During the first
years, the planet accumulates solids by rapid runaway accretion; this ``phase 1'' ends when the planet has severely depleted its feeding zone of
planetesimals. The accretion rates of gas and solids are nearly constant during most of the duration of the phase
2. The planet's growth accelerates toward the end of phase 2, and runaway accumulation of gas characterizes the phase
3. The simulation was stopped when accretion became so rapid that the model broke down. The endpoint was thus an artifact of the technique. As expected, the final mass of the planet is proportional to the density of planetesimals of the feeding zone, even if the core is initially of the same dimension.
The core of Jupiter predicted by these simulations is more massive than the current estimation. This is one of the issues which has to be explained by the core accretion models.  

The major problem of this theory is that the protoplanetary disk lifetime, few Myr, is shorter than the time needed for a core to grow. The growth timescale is proportional to $a^{3/2}M^{-1/2}_{\star}$, for that planets far away from the host star cannot be formed there \citep{2011exop.book..319D}. To solve this issue, \citet{1984Icar...60...29H} first proposed the mechanism of migration. In this way, planets can move from their formation position. Migration occurs when the torques exerted by the different parts of the disk interact with the planet. It reacts by adjusting its semi-major axis to compensate for the forces. There are primarily two types of migration: type I, the migration of low mass planets \citep[see][]{1980ApJ...241..425G, 2002ApJ...565.1257T}, and type II, where massive planets open a gap in the protoplanetary disk and migrate \citep{1986ApJ...309..846L}. This mechanism permits the core to accrete on different zones where it can find new planetesimals as the lengthy phase II is skipped \citep{2004A&A...417L..25A}.
      
 The Solar system can be at the same time good evidence for the core accretion model and against it. The outer Solar system is consistent with the trend of time scale/radius. The less massive planets could have been formed during the beginning of the dissipation of the gas. Also, Jupiter has the closest Solar composition, while the other giant planets are poorer in gas \citep{2007astro.ph..1485A}.

Against the model, there is the difficult explanation of the Neptune time scale. To avoid this problem, migration of the planet is proposed. Also problematic is to explain the small core of Jupiter, as we saw before, even if a core erosion may have occurred.

Concerning exoplanets, the correlation between the frequency of planets and the metallicity of the host seems to confirm the core accretion scenario, as it is more probable to form planets faster around high metallicity stars. On the other hand, one should consider that this statistical analysis is biased by the high number of close-in planets discovered so far, and it could be different for long-period planets.

To avoid the problems raised by the core accretion model and to explain exceptions found in the exoplanet population, especially for objects at great distances from the host, another scenario has been proposed, presented in the next Section. 
 
\subsection{Gravitational instability}

An alternative theory to the core-accretion model suggested especially to obviate the timescale problem, is the {\it gravitational instability} scenario \citep{1978M&P....18....5C, 1997Sci...276.1836B}.  Following this theory it is possible to explain the formation of massive planets, far from their host star. A condition for this scenario is that the disk is massive. Instability perturbations may occur in a disk if the Toomre stability condition falls: 
\begin{equation}
  Q = \dfrac{c_s\kappa_e}{\pi G\sigma} \sim \frac{M_{star}}{M_{d}}\frac{H}{r} < 1
\end{equation}
where $c_s$ is the speed of sound, $\kappa_e$ is the epicyclic frequency in one point of the disk, $\Sigma \sim M_{d}/r^2$ is the surface density, $H$ is the disk vertical scale height and $M_d$ is the mass of the disk within the radius $r$ \citep{1964ApJ...139.1217T}. If this occurs, the disk fragments into pieces, and from the clumps that they cause, future giant gas planets can form.

\begin{center}
	\begin{figure}[!htp]
	\centering
	\includegraphics[width=14cm]{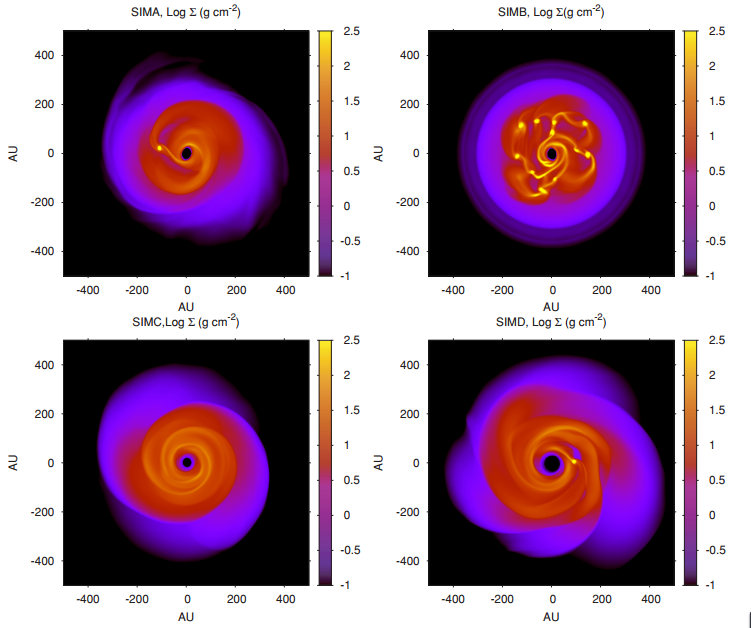}
	\caption{Figure taken from \citet{2009ApJ...695L..53B}. Snapshots of the surface density of the disk at the end of different simulations to predict gravitation instability formation. In three out of four cases, the gravitational instability bursts lead to fragmentation and formation of clumps.}
        \label{f:grav_ins}
	\end{figure}
	\end{center}

A steady-state disk becomes less stable at large radii \citep{2007astro.ph..1485A}. If the mass accretion rate increases, the radius decreases. Examples of simulations by \citet{2009ApJ...695L..53B}, performed to predict the formation of planets by gravitational instability are shown in Fig.~\ref{f:grav_ins}. Three out of four synthetic disks present clumps at the end of the simulation. 

Another parameter that plays an important role is the temperature of the disk. Lower-temperature disks may be more likely to be unstable. 
This mechanism is faster than the core-accretion one, taking less than 1 Myr. This is due to the fact that disk fragmentation is possible if the cooling time is shorter than the orbital period.

\begin{figure}
  \begin{center}
    \includegraphics[width=0.8\textwidth]{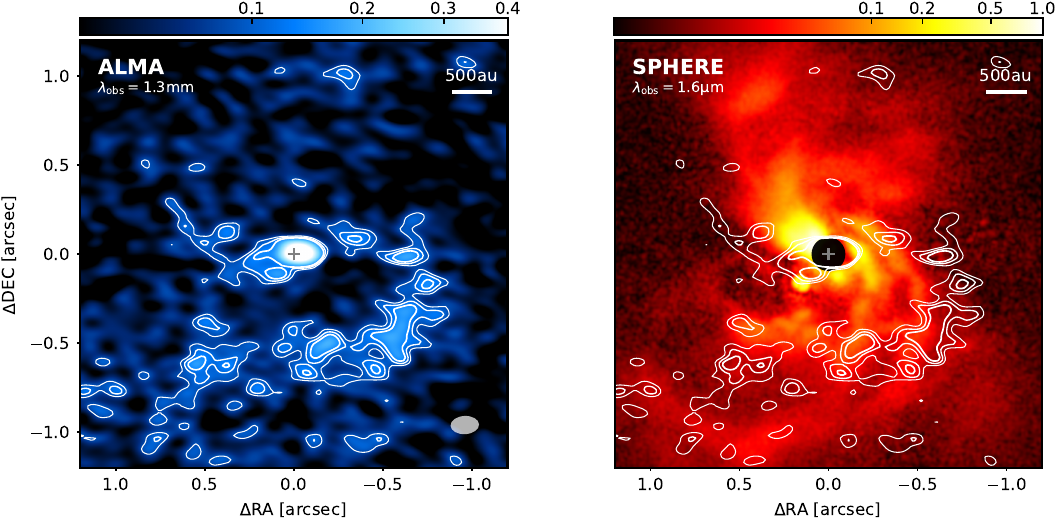}
  \end{center}
  \caption{First evidence of gravitational instability clumps with planetary masses. On the left, the ALMA band 6 continuum image of the protoplanetary disk around the FUor star V960 Mon, and on the right the SPHERE/IRDIS polarized light image. Contours of the ALMA continuum are overlaid on both images, corresponding to levels of 3, 4, and 5 times $\sigma$. The ALMA signal showing the clumps share the same location of the spiral arms around the star. Image from \citet{2023ApJ...952L..17W}. }
  \label{f:vmon}
\end{figure}

The first-time evidence of a spiral arm of a disk fragmenting into planet-forming clumps was found around the FUor star V960 Mon \citep{2023ApJ...952L..17W}. Comparing scattered light images of the massive disk around the star with SPHERE and ALMA data they discovered that unresolved clumps are detected in the sub-mm at the same location of the spiral arm seen in scattered light (see Fig.~\ref{f:vmon}).

This is the first direct detection of gravitational instability clumps in the planetary-mass regime. This direct detection of clumps within a protoplanetary disk represents an unprecedented achievement in the field. 
Studying FUor objects and their connection to gravitational instability is critical not only for understanding episodic accretion processes but also for deciphering the formation of gas giant planets and the controversy between GI and CA scenarios. In recent years, the latter has been favored due to a lack of observational evidence supporting GI. At the moment of observation, the majority of protoplanetary environments do not meet the mass requirements necessary to initiate GI, and the ordered, smooth structures commonly found in Class II disks appear to contradict the occurrence of large-scale instabilities.

The planets formed by gravitational instability are expected to be far away from the host star and to have great planetary radii. Also, differently from the core accretion model, no core is expected. 
This model predicts that planets have a high initial entropy. For this reason, we define ``hot-start'' the birth of planets in the gravitational instability scenario, and ``cold-start'' the one of the core accretion. This is crucial when we want to assume a mass for planets during the first stages of their life, as the ``hot-start'' planets are much brighter for a fixed mass. On the other hand, \citet{2013A&A...558A.113M} proposed that planets with massive cores, formed by core accretion, can also have high entropy during the first stages.   

This scenario has been proposed to explain some of the directly imaged systems as  HR\,8799 \citep{2010Natur.468.1080M}, or planets around multiple systems \citep{2008ApJ...681..375K, 2010ApJ...708.1585K,2010ApJ...710.1375K}. Also, it could explain the fact that Jupiter has a smaller core than expected from core accretion. Finally, planets formed by core accretion have a limit on the mass, depending on the critical mass reached by the core. On the other hand, gravitation instability can explain massive planets formation. \citet{2009ApJ...695L..53B} proposed to unify these two theories that can coexist, by demonstrating that the core accretion can operate in the internal part of the disk ($\sim$ 100 au) while gravitational instability clumps can form planets in the outer region. 

The direct imaging method will help us understand the formation of planets in wide orbits, which are not explored by the other indirect techniques. Also, measuring the luminosity of planets during the first stages of their life will help us understand which is the scenario that better describes the formation of each individual planetary system. Coupling the estimated mass from ``hot-start'' and ``cold-start'' models and dynamical information on the mass will be fundamental to comprehending how giant planets form.   

\section{Protoplanets}
\label{s:proto}

\begin{center}
	\begin{figure}[!htp]
	\centering
	\includegraphics[width=\textwidth]{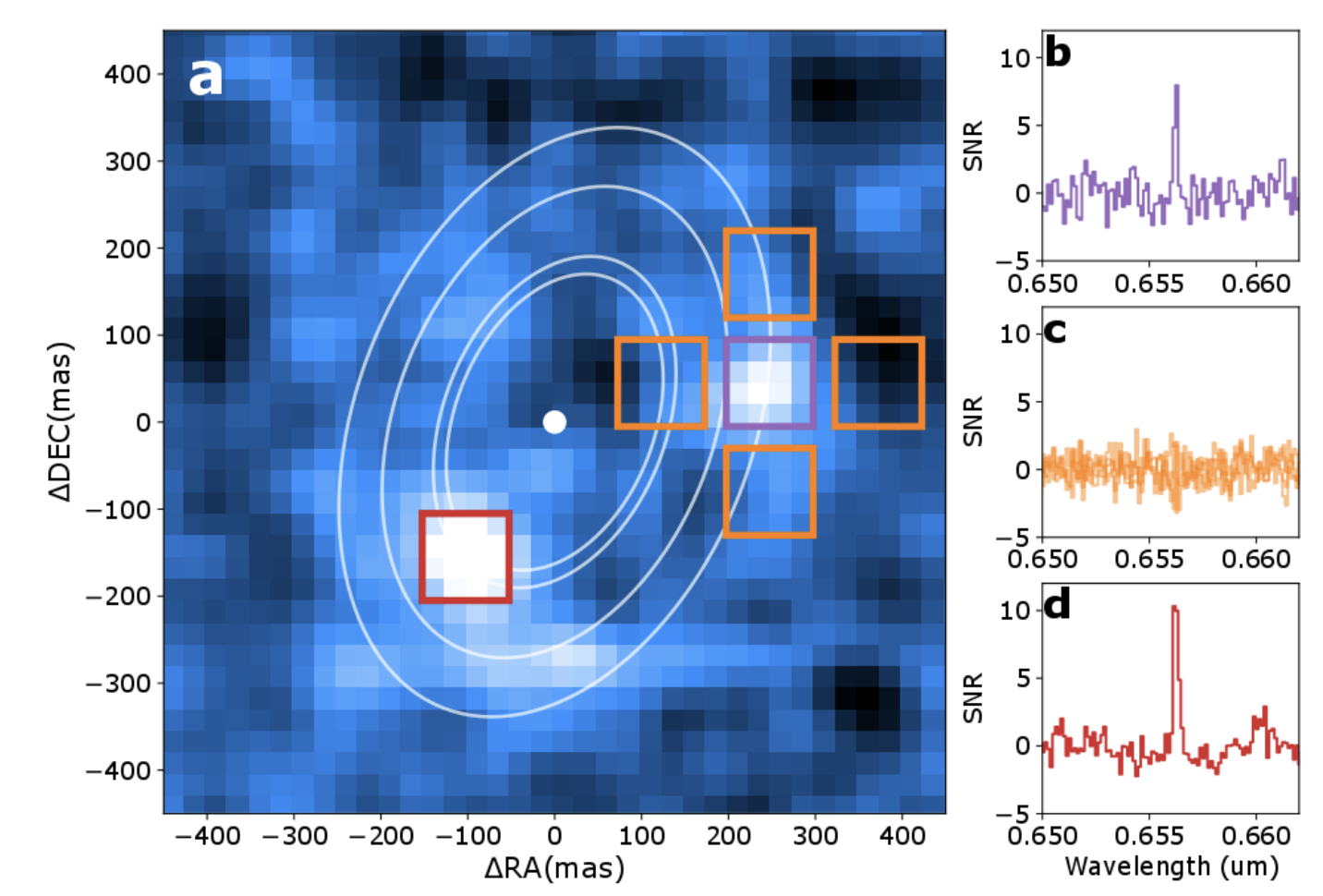}
	\caption{Figure and caption taken from \citet{2019NatAs...3..749H}. Accreting protoplanets around the system of PDS 70. The H$\alpha$ detection map with an overlay of the contours of the
orbital radii and a white dot in the center that marks the position of the star. The contours for PDS 70 c
are the minimum and maximum orbital radii found for the different wavelength observations. For both
objects the square apertures that were used for the photometry are shown, with the red aperture for
PDS 70 b and the purple aperture for PDS 70 c. b, c, d, The corresponding spectra divided by their
standard deviation are on the right and centered around the H$\alpha$ line position. The four apertures in
orange indicate reference areas that are used to compare with PDS 70 c. The orange reference spectra on the right do not show any spectral feature, while both PDS 70 b and c clearly show H$\alpha$ in emission.}
        \label{f:haffert}
	\end{figure}
	\end{center}

Understanding the formation of planets is of paramount importance in astrophysics, with implications for our understanding of the origin and diversity of planetary systems \citep[see, e.g.,][]{ 2012ApJ...745..174S}. Previous breakthroughs in HCI have highlighted the potential of detecting protoplanets, or still forming planets, through the H$\alpha$ emission, which serves as an indicator of accretion onto compact bodies \citep[see, e.g.,][]{2012A&A...548A..56R}. Notable successes include the detection of a close stellar companion around HD142527 \citep{2014ApJ...781L..30C, 2019A&A...622A..96C} and the strongest detection to date: two accreting protoplanets around the star PDS 70, which hosts a gapped protoplanetary disk \citep[][and Fig.~\ref{f:haffert}]{keppler,2018ApJ...863L...8W,2019NatAs...3..749H}.

Approximately $\sim$30\% of Herbig Ae/Be disks should host giant planets of $\sim$0.1 to 10 \MJup \citep{2015A&A...582L..10K}, and should be classified as ``transitional disks'' (TDs), which are peculiar disks with cavities, gaps, and spiral structures that can be induced by the presence of such a companion \citep[e.g.,][]{2011ApJ...738..131D,2023A&A...670L...1S}. Supporting this theory, hydrodynamical simulations predicted the presence of substellar companions around some observed TDs \citep[e.g.,][]{2015ApJ...812L..32D}, but the putative objects have not been detected yet. Moreover, the planet's radiative feedback (i.e., intrinsic planet luminosity as a consequence of planet accretion) is an important ingredient that affects the evolution, morphology, and emission of the disk \citep{2015ApJ...806..253M}.

To confirm a protoplanet detection remains a difficult task, in fact in the past several candidates turned out to be disk features rather than forming planets. The first protoplanet candidate was announced in \citet{2015Natur.527..342S}, around the young star LkCa 15. They found H$\alpha$ emission from a source close to the inner ring of the protoplanetary disk. The discovery was later on revisited by \citet{2016ApJ...828L..17T}, where the feature was presented as a portion of the disk.  Other candidates are not confirmed, for example, the ones around the star HD 100546 \citep{2013ApJ...766L...1Q,2015ApJ...814L..27C}. Two planets have been detected, but following studies did not confirm their orbital motion and were not detected in H$\alpha$ \citep{2017AJ....153..244R,2018A&A...619A.160S}.

An emblematic discovery is the protoplanet around AB Aur \citep{2022NatAs...6..751C}. The companion is located at a wide projected separation from the star (93 au) and it shows emission in H$\alpha$. The system shows spectacular spiral arms, an indicator of the presence of perturbing companions. Similarly, an embedded young planet has been found around the star MWC 758 \citep{2023NatAs...7.1208W}. This planet, at a projected separation of 100 au, is compatible with the shaping of the spiral arms around the system. No signal is recovered in H$\alpha$ at the location of the planet \citep{2019A&A...622A.156C, 2020A&A...633A.119Z}.

Exploring the connection between the presence of (sub)-stellar companions and the structures observed in many protoplanetary disks, such as gaps, spirals, and rings, remains a significant challenge. The limited visibility of many newborn planets, still embedded in their disks and entangled with scattered light from the circumstellar environment, could explain the absence of detection in most cases. Factors such as the less favorable Strehl ratio in visible light and disk extinction at visible wavelengths can hinder their detection.

Surveys dedicated to looking for still accreting planets in the H$\alpha$ line have unfortunately been very challenging in the past \citep[e.g.,][]{2019A&A...622A.156C, 2020A&A...633A.119Z,2022A&A...668A.138H}. The Strehl ratio is less favorable in visible light, and the disk extinction in the visible can play a significant role. On the other hand, in the thermal infrared range, the contribution of circumplanetary material can enhance the signal at the planet's location  \citep[see, e.g.,][]{2016ApJ...832..193Z}. Furthermore, accreting protoplanets are expected to host circumplanetary disks \citep[e.g.,][]{2018ApJ...866...84A}. Therefore, younger planets have higher chances of detection at longer wavelengths due to the presence of circumplanetary disks and their high entropy during the first Myr of their formation. New instruments such as VLT/ERIS, JWST/NIRCAM, or, in the future, eELT/METIS will play a crucial role in detecting still-forming planets at longer wavelengths.

An alternative technique to look for protoplanets still embedded in their parent disk is the ``disk kinematics'' technique \citep{2015ApJ...811L...5P,2018MNRAS.480L..12P}, which allows to pinpoint the presence and locations of protoplanets via the footprint they stamp on the velocities of the gas in a disk. Recently we saw the first indirect detection using disk kinematics of a protoplanet \citep[HD 163296 b;][]{2018ApJ...860L..13P} and a protoplanet inside a gap \citep[HD 97048 b;][]{2019NatAs...3.1109P}. This technique is now widely used in the international community and currently drives some of the most ambitious proposals for ALMA observations.

\section{Brown dwarfs and exoplanets}
\label{s:atm}
Similar properties of the giant planets of the solar system are found in the so-called ``substellar objects'': brown dwarfs (BDs) and planets. This class is composed of objects that are not massive enough to host nuclear fusion of hydrogen in the core. The upper limit of the mass for a substellar object is then $\sim~0.08$ M$_{\odot}$ (84 \MJup). Objects in the mass range 11.0--16.3 \MJup can fuse deuterium \citep{2011ApJ...727...57S}, and those above $\sim$ 65 \MJup can fuse lithium \citep{2006asup.book....1L}. The boundary between a brown dwarf and a giant planet is still in debate. More studies will permit us to know whether the mechanisms of formation and/or the physics of the interior are different. Generally, the limit is imposed by the mass of the object for simplicity, but it should be given by the formation mechanism.

\begin{center}
	\begin{figure}[!htp]
	\centering
	\includegraphics[width=\textwidth]{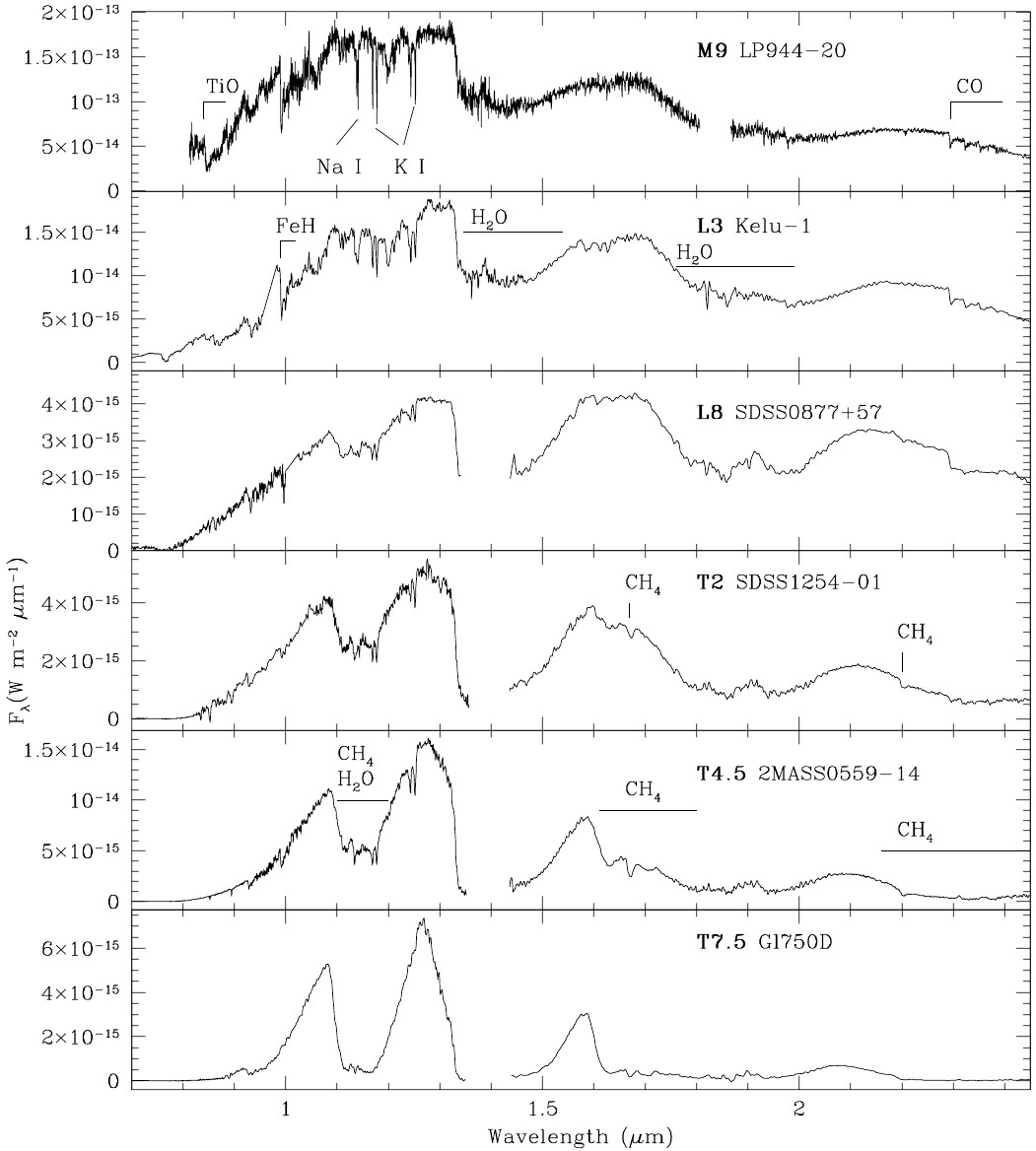}
	\caption{Figure and caption taken from \citet{2014PASA...31...43B}, with references for the data therein. Spectra of ultracool dwarfs from M9 to T7.5. The species responsible for the main absorption features are indicated. }
        \label{f:sp_seq}
	\end{figure}
	\end{center}

\begin{center}
	\begin{figure}[!htp]
	\centering
	\includegraphics[width=0.8\textwidth]{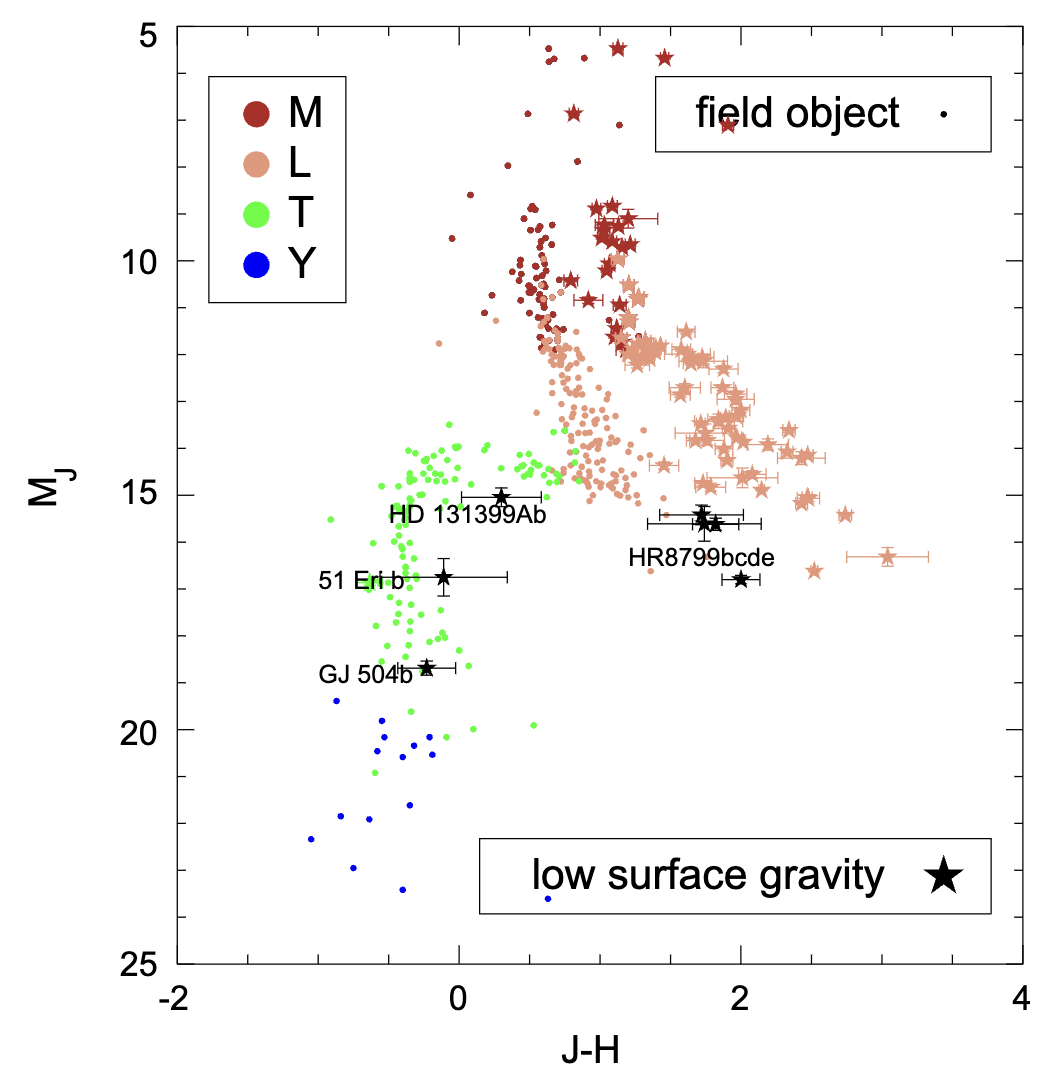}
	\caption{Figure taken from \citet{2017AstRv..13....1B}, with references for the data therein. Color magnitude diagram ($J-H$ against absolute magnitude in $J$, $M_J$) for
late-type dwarfs from spectral type M to Y. Some direct imaging companions are represented as black stars.}
        \label{f:cc_seq}
	\end{figure}
	\end{center}

The class of substellar objects includes spectral types from late M to Y. The spectral types L and T were introduced by \citet{1999ApJ...519..802K}, for objects cooler than M-dwarfs. Then, \citet{2011ApJ...743...50C} introduced the class of Y-type dwarfs, the coolest objects ever observed. The main features of substellar objects of the spectral sequence from M to T are shown in Fig.~\ref{f:sp_seq}. The main properties of the spectral classes are:
\begin{itemize}
\item {\bf M-dwarfs}. This spectral type is the link between stellar objects and BDs. The latest types of the M sequence include young objects which are considered BDs. The spectrum is characterized by the presence of TiO and VO. TiO bands increase in strength
up to spectral type M6, and VO becomes strong in the
latest types \citep{2014PASA...31...43B}.
Broad
absorptions due to H$_{2}$O are found around 1.4 and 1.9 \mic,
especially in later spectral types. Other molecules with strong absorption are FeH, CrH, and MgH.
\item {\bf L-dwarfs}. In this type of dwarfs, the TiO and the vanadium monoxide are increasingly disappearing. At their places, the bands of CrH, FeH, and alkaline metals appear in the optical. In the near-infrared, the bands of CO, CrH, and FeH are strong till the mid-L and then become weaker. The sequence of the L-dwarfs is redder than the others in the NIR mag/color diagram (see Fig.~\ref{f:cc_seq}). This is due to the dusty clouds of species such as enstatite, forsterite, spinel, and solid iron which condense in the upper layers of the atmosphere \citep[see, e.g., ][]{2001ApJ...556..357A}.

\item {\bf T-dwarfs}. The T-dwarf class is characterized by the presence 
of methane (CH$_4$) absorption features in the near-IR
region (1-2.5 $\mu$m) and water absorption bands. The methane absorptions at 1.6 and 2.2 \mic ($H$ and $K$ bands) cause the blue shift of the sequence in the mag/color diagram (see Fig.~\ref{f:cc_seq}). They have a closer resemblance to the solar system's giant planets. \citet{2002ApJ...564..421B} estimate that, at birth, Jupiter had a T$_{eff}$ near 600-1000 K and the appearance of a T-dwarf. 

\item {\bf Y-dwarfs}. These objects lack water absorption bands because the H$_2$O condensates with temperatures around 500 K and it is not present in their clouds \citep{2001ApJ...556..872A,2003ApJ...596..587B}.
\end{itemize} 

\begin{center}
	\begin{figure}[!htp]
	\centering
	\includegraphics[width=0.7\textwidth]{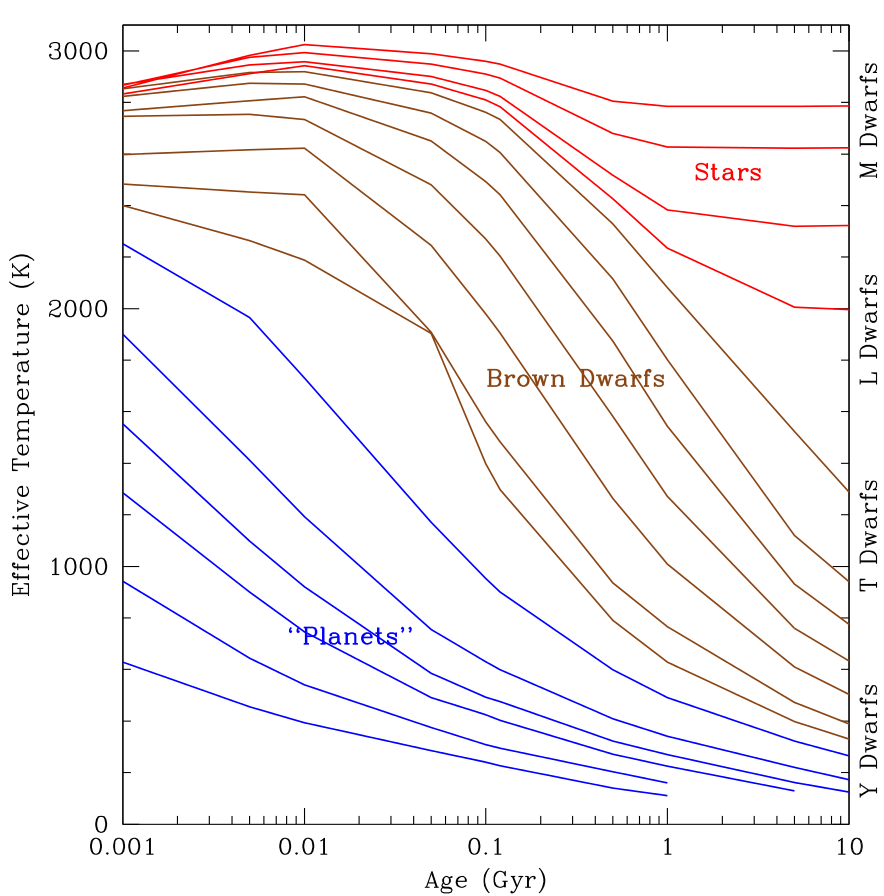}
	\caption{Figure taken from \citet{2014PASA...31...43B}. Evolution of effective temperature for substellar objects. The range of mass represented goes from 0.0005 to 0.1 M$_{\odot}$  based on the models of \citet{2003A&A...402..701B}. The
red tracks are for stars with masses above the hydrogen-burning
limit. The magenta tracks are for brown dwarfs and the blue
tracks are for objects below the deuterium burning limit (planets
or sub-brown dwarfs) The tracks plotted from top to bottom are
masses of (Stars: 0.1, 0.09, 0.08, 0.075 M$_{\odot}$) (Brown Dwarfs: 0.07,
0.06, 0.05, 0.04, 0.03, 0.02, 0.015 M$_{\odot}$), (Planets: 0.01, 0.005, 0.003,
0.002, 0.001, 0.0005 M$_{\odot}$).}
        \label{f:temp_age}
	\end{figure}
	\end{center}

Substellar objects are known to evolve with time more rapidly than low-mass stars. This is a fundamental property of these objects for the direct imaging technique, as in the first stages of their life they are much brighter and hotter, making their detection by HCI much easier. Since they do not have any significant internal source of energy, planets, and BDs cool down gradually with time. The evolution in time of the temperature for different masses is shown in Fig.~\ref{f:temp_age}. Brown dwarfs have an initial temperature of $\sim$2500 K during the first 10 Myr, then they cool rapidly and the temperature can drop to 500 K after 10 Gyr. On the other hand, planets cool down since the beginning, reaching a final temperature of $\sim$ 200 K after 1 Gyr.  In the first stages, substellar objects are bright, large in radius, and usually fully adiabatic (as Jupiter). 

\begin{center}
	\begin{figure}[!htp]
	\centering
	\includegraphics[width=0.7\textwidth]{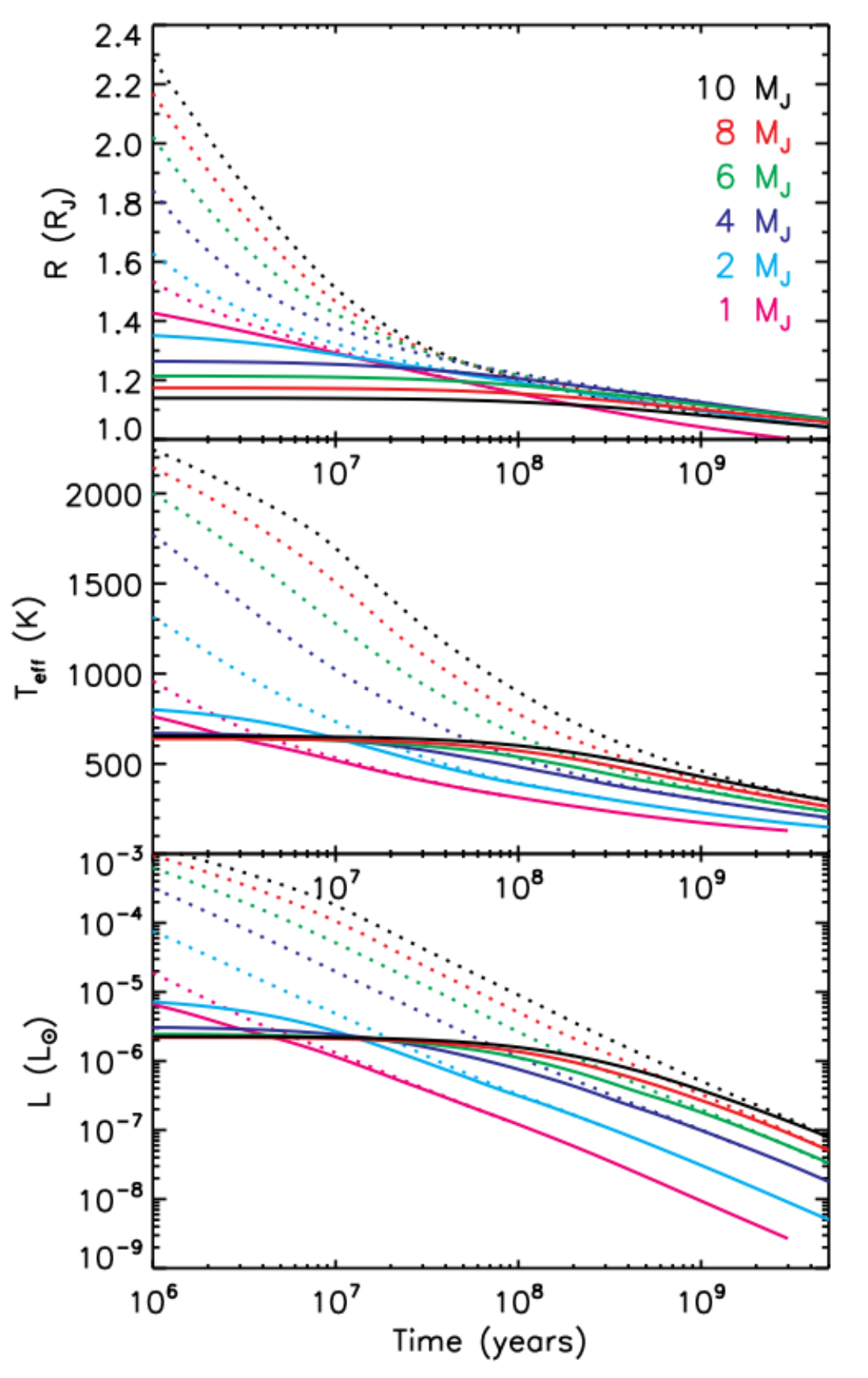}
	\caption{Figure taken from \citet{2007ApJ...655..541M}. Evolution of the parameters of radius, $R$, temperature, T$_{eff}$, and luminosity, $L$, for planets with different masses. ``Hot-start'' models are represented in dotted lines. The parameters for the two models converge for older ages. The moment when the planet loses the information on its birth depends on its mass.}
        \label{f:hot_cold}
	\end{figure}
	\end{center}

The luminosity of substellar objects generally decreases with a power law (roughly $L \sim 1/t$). If we combine the information on the luminosity with the estimated age of the system, we can derive a mass for the companion. This value is only an estimated quantity from models if not coupled with dynamical information (from radial velocities, for example). Especially at young ages, models are not fully tested. \citet{2007ApJ...655..541M} proposed models where the initial conditions were not arbitrary but assumed by the core accretion scenario: the ``cold-start'' models. The two different formation scenarios seen in the previous Section~\ref{s:form}, lead to different predictions on the luminosity of planets at the first stages. Planets formed by gravitational instability are expected to have large radii and high temperatures. The entropy of these objects is high. In the core accretion scenario, the birth of planets is more gentle, they have a low entropy, small radii and low T$_{eff}$. The evolution of the three parameters is shown in Fig.~\ref{f:hot_cold} for both models. The plots show that planets keep the information of their formation just in the first Myr of their life, for old objects the two models are coincident. 
\citet{2012ApJ...745..174S} proposed a third scenario, called ``warm-start'', which is in between the two former models. Direct imaging campaigns, possibly coupled to astrometric or RV information, are providing luminosities of newborn planets that will help to constrain and test these evolutionary tracks and better understand the mechanism of formation and evolution of substellar objects.   

\section{Conclusions and future perspectives}
\label{s:conc}
Direct imaging stands out as the exclusive technique capable of disentangling the light emitted by extrasolar planets from that of their parent stars. This method offers a unique opportunity to delve into the outer regions of planetary systems, for separations greater than 10 astronomical units from the host star—an exploration not feasible with indirect detection techniques. A key advantage of direct imaging lies in its ability to capture the spectra of these distant worlds, providing valuable insights into their composition and chemistry.

Moreover, direct imaging, when coupled with sufficient time coverage, facilitates the comprehensive determination of orbital parameters for the detected objects. This precision is particularly notable for close-in planets such as $\beta$ Pic b. High-contrast imaging provides spectra from sub-stellar companions, a powerful instrument to investigate the properties of planetary atmospheres, cloud coverage, chemical abundances, etc.  

The primary targets of high-contrast imaging are very young systems, as planets are self-luminous and bright during the first stages of their life. Direct imaging provides then a whole picture of a newly formed system, giving the opportunity to investigate which is the preferred mechanism of planet formation, a debate that still remains open. 

In particular, the direct imaging captured photons of still-forming planets, or ``protoplanets'', using mass accretion indicators as the H$\alpha$ line emission. Current detections are only a few due to the difficulty of disentangling the signal of a forming planet from a disk feature. The strongest detection to date is two accreting protoplanets around the young star PDS 70, which also hosts a gapped protoplanetary disk. Future detections will help us understand the mass formation rate of substellar companions and the location where planets form in a disk.  

The future of direct imaging is promising. Instruments on board the James Webb Space Telescope are delivering their first spectacular results. JWST will permit thermal infrared (3--20 \mic) observations of exoplanets, which are heavily affected by the thermal background for ground-based telescopes. From the ground, the instrument VLT/ERIS started its operations in 2022. ERIS includes a near-infrared camera that operates with the APP and vortex coronagraphs and is suitable for detecting exoplanets even around faint stars. The instrument can use a laser guide star to perform the observations.

The first light for the European Extremely Large Telescope (eELT) is foreseen for the year 2028. One of the first light instruments is METIS, a mid-infrared imager and spectrograph that will be equipped with coronagraphs to perform high-contrast imaging. After the first generation instruments, a dedicated planet finder, the planetary camera and spectrograph (PCS) instrument is foreseen to be installed in the eELT \citep{2021Msngr.182...38K}. This instrument is specifically designed to image even Earth-size planets in
the neighborhood of the Sun. Instruments from space and the ground will be able soon to directly detect planets with the mass of our Earth, with tremendous implications for the study of planetary systems, their formation, and evolution.

\begin{ack}[Acknowledgments]
TThe author acknowledges support from ANID -- Millennium Science Initiative Program -- Center Code NCN2021\_080.
\end{ack}

\seealso{The Protostars and Planet VII Chapter ``Direct Imaging and Spectroscopy of Extrasolar Planets'' by T. Currie et al. and ``An Introduction to High Contrast Differential Imaging of Exoplanets and
Disks'' by K. Follette.}

\bibliographystyle{Harvard}
\bibliography{astro-ph}

\end{document}